# Characterization of Fault Roughness at Various Scales: Implications of Three-Dimensional High Resolution Topography Measurements

Thibault Candela [1], François Renard [1, 2], Michel Bouchon [3], David Marsan[4], Jean Schmittbuhl[5], and Christophe Voisin [3]

(1) University Joseph Fourier Grenoble I, Laboratoire de Géodynamique des Chaînes Alpines, CNRS, BP 53, 38041 Grenoble, France (Thibault.Candela@bvra.e.ujf-grenoble.fr, françois.renard@ujf-grenoble.fr).

(2) Physics of Geological Processes, University of Oslo, Oslo, Norway.

(3) University Joseph Fourier Grenoble I, Laboratoire de Géophysique Interne et Tectonophysique, CNRS, Grenoble, France (christophe.voisin@ujf-grenoble.fr, michel.bouchon@ujf-grenoble.fr).

(4) University of Savoie, Laboratoire de Géophysique Interne et Tectonophysique, CNRS, Le Bourget du Lac, France (david.marsan@univ-savoie.fr)

(5) UMR 7516, Institut de Physique du Globe de Strasbourg, Strasbourg, France.

Short title: Roughness of fault surfaces

*Abstract* – Accurate description of the topography of active faults surfaces represents an important geophysical issue because this topography is strongly related to the stress distribution along fault planes, and therefore to processes implicated in earthquake nucleation, propagation, and arrest.

Up to know, due to technical limitations, studies of natural fault roughness either performed using laboratory or field profilometers, were obtained mainly from 1D profiles. With the recent development of Light Detection And Ranging (LIDAR) apparatus, it is now possible to measure accurately the 3D topography of rough surfaces with a comparable resolution in all directions, both at field and laboratory scales. In the present study, we have investigated the scaling properties including possible anisotropy properties of several outcrops of two natural fault surfaces (Vuache strike-slip fault, France, and Magnola normal fault, Italy) in limestones. At the field scale, digital elevation models of the fault roughness were obtained over surfaces of 0.25 $m^2$ to 600 $m^2$ with a height resolution ranging from 0.5 mm to 20 mm. At the laboratory scale, the 3D geometry was measured on two slip planes, using a laser profilometer with a spatial resolution of 20 μm and a height resolution less than 1 μm.

Several signal processing tools exist for analyzing the statistical properties of rough surfaces with self-affine properties. Among them we used six signal processing techniques: (i) the root-mean-square correlation (RMS), (ii) the maximum-minimum height difference (MM), (iii) the correlation function (COR), (iv) the RMS correlation function (RMS-COR), (v) the Fourier power spectrum (FPS), and (vi) the wavelet power spectrum (WPS). To investigate quantitatively the reliability and accuracy of the different statistical methods, synthetic self-affine surfaces were generated with azimuthal variation of the scaling exponent, similar to what is observed for natural fault surfaces. The accuracy of the signal processing techniques is assessed in terms of the difference between the "input" self-affine exponent used for the synthetic construction and the "output" exponent recovered by those different methods. Two

48  kinds of biases have been identified: artifacts inherent to data acquisition and intrinsic errors
49  of the methods themselves. In the latter case, the statistical results of our parametric study
50  provide a quantitative estimate of the dependence of the accuracy with system size and
51  directional morphological anisotropy.
52  Finally, we used the most reliable techniques (RMS-COR, FPS, WPS). For both field and
53  laboratory data, the topography perpendicular to the slip direction displays a similar scaling
54  exponent $H_\perp = 0.8$. However, our analysis indicates that for the Magnola fault surface the
55  scaling roughness exponent parallel to the mechanical striation is identical at large and small
56  scales $H_{//} = 0.6 - 0.7$ whereas for the Vuache fault surface it is characterized by two different
57  self-affine regimes at small and large scales. We interpret this cross-over length scale as a
58  witness of different mechanical processes responsible for the creation of fault topography at
59  different spatial scales.

60

61

62

63

64

65

66

69

## 1. Introduction

Knowledge of the detailed fault geometry is essential to understand some major processes involved in faulting such as grain comminution or asperities abrasion during slip, geometrical heterogeneity of the stress field that controls earthquake nucleation (CAMPILLO *et al.*, 2001; VOISIN *et al.*, 2002a), rupture propagation, and arrest (VOISIN *et al.*, 2002b). Establishing correlations between geometrical properties of fault roughness (POWER *et al.*, 1987, 1988; POWER and TULLIS, 1991; SCHMITTBUHL *et al.*, 1993; LEE and BRUHN, 1996; POWER and DURHAM, 1997; RENARD *et al.*, 2006; SAGY *et al.*, 2007), seismic behavior of faults (OKUBO and AKI, 1987; PARSON, 2008), frictional strength and critical slip distance (SCHOLZ, 2002), wear processes during fault zone evolution (POWER *et al.*, 1988) represents a fundamental issue to understand seismic faulting.

At the laboratory scale, AMITRANO and SCHMITTBUHL (2002) highlight a complex coupling between fault gouge generation and fault roughness development. At larger scale, asperities control the slip distribution of earthquake (PEYRAT *et al.*, 2004). Indeed asperities on active fault planes concentrate the stress (MARSAN, 2006; SCHMITTBUHL *et al.*, 2006) and therefore may control earthquake nucleation (LAY *et al.*, 1982; SCHOLZ, 2002) and the propagation of the rupture to its ultimate arrest (AKI, 1984). High resolution relocations of earthquakes using the multiplet technique have shown streaks of earthquake along several faults in California (RUBIN *et al.*, 1999). This pattern has been interpreted as resulting from the presence of an organized large scale roughness (asperities) resisting slip (SCHAFF *et al.*, 2002).

Despite recent progress in seismology, the imaging of fault planes over a large range of scales at depth is not yet available. A quasi-unique access to high resolution description of the fault plane comes from exhumed fault scarp observations. This requires, of course, that the main morphological patterns of faults mapped at the surface of the earth persist at least across the seismogenic zone. Due to technical limitations, prior comparative studies of natural fault roughness were mainly based on 1D profilometry (POWER *et al.*, 1987, 1988; POWER and TULLIS, 1991; SCHMITTBUHL *et al.*, 1993; LEE and BRUHN, 1996; POWER and DURHAM, 1997). These 1D measurements have shown that fault roughness can be characterized by a scale invariance property described by a self-affine geometry (see section 2 for the definition of self-affinity) with a roughness scaling exponent close to 0.8 for profiles oriented in a direction perpendicular to the striations observed on the fault plane. Such striations are aligned in the direction of slip. The value of 0.8 is similar to what was measured for tensile cracks (POWER *et al.*, 1987; SCHMITTBUHL *et al.*, 1995b; BOUCHAUD, 1997). Moreover, the influence of slip was also quantified: the fault surfaces have smaller roughness amplitude along the slip direction than perpendicular to it (POWER *et al.*, 1988; POWER and TULLIS, 1991; LEE and BRUHN, 1996; POWER and DURHAM, 1997). The compiled fault roughness statistics of several studies (POWER and TULLIS, 1991; LEE and BRUHN, 1996; BEN-ZION and SAMMIS, 2003) suggest a change in scaling properties between large and short length scales. However, considering the noise in their data, these authors underlined that it was not possible to decipher whether this variation was related to small-scale surface weathering of the fault scarp or to the faulting process itself.

With the recent development of high resolution distancemeters, it is now possible to use accurate statistical approaches to quantify fault roughness. Indeed, portable 3D laser scanners (also called LiDAR, Light Detection And Ranging) allow mapping fault surface outcrops over scales of 0.5 mm to several tens of meters. The accuracy of the measurements enables a reliable quantification of the data. RENARD *et al.* (2006) and SAGY *et al.* (2007) demonstrated precisely the scaling invariance and anisotropy properties of fault topography using ground based LIDAR and laboratory profilometers. In these previous studies, statistical analysis of fault roughness was carried out with a single signal processing tool. However,

- 5 -

SCHMITTBUHL *et al.* (1995a, 1995b) recommend the simultaneous use of different methods in order to appreciate the confidence in the measured scaling invariance properties.

In the present study, we use new roughness data to extend the type of measurements made by RENARD *et al.* (2006) and SAGY *et al.* (2007) and propose a parametric study of the statistical results of fault topography, using multiple signal processing tools. In order to investigate the reliability and accuracy of the different signal processing methods, synthetic self-affine surfaces were generated with azimuthal variation of the scaling exponent. These synthetic rough surfaces are completely characterized by two different self-affine exponents prescribed in perpendicular directions. When comparing these synthetic surfaces to natural faults, one should keep in mind that any self-affine model can only describe a real surface to a finite degree of accuracy, and only within a finite range of scales. After this systematic study, we used the most reliable and accurate techniques to investigate the scaling properties and anisotropy of several outcrops of two natural fault surfaces that have been measured using 3D laser scanners in the field and a laser profilometer in the laboratory.

This paper is organized as follows. In Section 2, following a brief explanatory discussion of the self-affine notion, the generation process of synthetic self-affine surfaces with a directional morphological anisotropy is described. In Section 3, statistical methods to define the self-affine properties are reviewed. Section 4 is devoted to the systematic study of the accuracy of the methods. Section 5 is devoted to the acquisition of data on natural fault. In Section 6, analysis of the roughness, covering six orders of magnitude of length scales, is performed on several outcrops of two natural faults. Finally, in Section 7, we conclude by linking our findings on the statistical properties of natural fault topography to the results of earlier studies, with the ultimate goal of developing a more mutually consistent description of fault asperities geometry.

## 2. Generation of self-affine surfaces

*2.1 Self-affinity in 1D and 2D*

Surface roughness analysis provides a statistical characterization of a surface which is simpler and easier to use than a complete deterministic description. In geophysics, BROWN and SCHOLZ (1985) and POWER *et al.* (1987) demonstrated the self-similar property of natural fault surfaces at field scale. Coming from statistical physics, a more general scaling transformation has been proposed: self-affinity (MANDELBROT, 1985; MANDELBROT, 1986; VOSS, 1985) that was successfully used for the quantitative description of fault roughness (SCHMITTBUHL *et al.*, 1993, RENARD *et al.*, 2006).

A self-affine 1D profile remains unchanged under the scaling transformation $\delta x \to \lambda\, \delta x$, $\delta z \to \lambda^H \delta z$ for 1D profiles (Figure 1) extracted from a surface (MEAKIN, 1998). Here, $\delta x$ is the coordinate along the profile and $\delta z$ is the roughness amplitude. For a self-affine profile, the scaling exponent $H$, also called Hurst exponent, lies in the range $0 \leq H \leq 1$. Accordingly, self-affinity implies that a profile appears less rough as the scale increases. In other words, if a profile is self-affine, a magnified portion of it will appear statistically identical to the entire profile if different magnifications are used in the x and z-directions (Figure 1).

For 2D surfaces, this self-affinity property can be described for sets of 1D parallel profiles extracted from the surface. Moreover, if the surface is striated along some given orientation, anisotropic scaling behavior can emerge if $H$ varies for different directions in the plane of the surface. An anisotropic self-affine surface $Z(x, y)$ with coordinates $(x, y)$ obeys the property: $Z(\lambda^{1/H_{//}} x, \lambda^{1/H_\perp} y) = \lambda\, Z(x, y)$, where $\lambda$ is a positive dilation factor, $H_{//}$ and $H_\perp$ are the Hurst exponents, comprised between 0 and 1, in two perpendicular directions of the surface. $H_{//}$ is defined along a direction parallel to the main striations, and $H_\perp$ is defined along a direction perpendicular to the striation (Figure 1b).

*2.2 Synthetic anisotropic self-affine surfaces*

To calculate synthetic fault surfaces (Figure 1b), we used a Fourier based method to simulate a fractional Brownian motion on a 2D grid (STEIN, 2002), where an anisotropy matrix $E = \begin{pmatrix} 1/H_{//} & 0 \\ 0 & 1/H_{\perp} \end{pmatrix}$ was introduced when calculating the 2D Gaussian random field. The eigenvalues of this matrix correspond to the inverse of the two roughness exponents $H_{//}$, and $H_{\perp}$ that characterize the self-affine properties of the generated surface in two perpendicular directions (BIERME *et al.* 2007, 2008). The code to generate an anisotropic 2D self-affine surface, written in Matlab©, is given in the appendix A and can be run easily on a desktop computer.

In the following sections, we decompose the signal processing analysis of rough surfaces in two stages. Firstly, we present the six signal processing tools used to estimate the self-affine property of an isotropic surface with a single Hurst exponent (Figure 1a), as observed for example for fresh mode I brittle fractures in rocks (POWER *et al.*, 1987; SCHMITTBUHL *et al.*, 1995b; BOUCHAUD, 1997). For this, we have synthesized several isotropic surfaces with an exponent in the range [0.1 - 0.9] and grid sizes in geometrical series: 129 × 129 points, 513 × 513 points, 2049 × 2049 points. Secondly, we analyse synthetic anisotropic surfaces (Figure 1b) with $H_{//}$ in the range [0.7 - 0.9] and $H_{\perp}$ in the range [0.4 - 0.9].

## *3. Statistical signal processing methods*

We have used six different methods that characterize the amplitude of the roughness at various spatial wavelengths. All these methods, presented in the following sub-sections, are based on the analysis of 1D profiles (Figure 1c) that are extracted from the 2D Digital Elevation Model (DEM) of 2D surfaces (Figure 1a, b). For each surface, a set of 1D parallel profiles in a specific direction are extracted, detrended and then analyzed. Then, the properties are averaged over all the 1D profiles to characterize the 2D surface in the chosen direction. We have repeated such analyses for profiles extracted in several directions, following a 360° rotation, allowing then to determine the azimuthal dependence of the statistical properties of the surface (RENARD *et al*., 2006) that could characterize a morphological anisotropy.

For the application to natural fault surfaces, we also tested how the noise in the data and the presence of missing points could affect the estimation of fault surface. Indeed, the raw scanner data consist of clouds of points, with x, y, and z coordinates, sampled more or less regularly. Sometimes, data are missing (vegetation on the fault plane, low reflectivity of the scanner light beam), and the surface is incomplete. An interpolation is then necessary, which induce a bias in the estimation of scaling exponents that need to be estimated too.

*3.1 Root-mean-square correlation (RMS) and maximum-minimum height difference (MM) methods*

Let consider a 1D profile $L(x)$. This profile is divided into windows of width $\delta x$ and indexed by the position of the first point $x_0$ of the band. The standard deviation $\sigma(\delta x)$ of the height $L(x)$ and the height difference $h(\delta x)$ between the maximum and minimum height are computed for each band, and then averaged over all the possible bands of fixed width $\delta x$ spanning the profile, by varying the origin $x_0$. We then obtain $\langle \sigma(\delta x) \rangle$ and $\langle h(\delta x) \rangle$, where both quantities follow a power law for a self-affine profile: $\langle \sigma(\delta x) \rangle \propto \delta x^H$ and $\langle h(\delta x) \rangle \propto \delta x^H$ (SCHMITTBUHL *et al*. 1995a).

Note that these techniques are useful when *H* is not too close from 0 or 1, where a significant error can be measured (see Figures 3a, b – 4a, b). Usually, levelling off of $\sigma(\delta x)$ at small $\delta x$ values is due to the noise in the data (see Figure 7c, d), and leveling-off at large $\delta x$ is due to the finite size of the profile.

218   *3.2 Height-height correlation function (COR) method*

219   For a signal $L(x)$, we consider the height-height correlation function defined by

220   $C(\Delta x) = \left[\left\langle (L(x) - L(x + \Delta x))^2 \right\rangle\right]^{1/2}$, which estimates the average height-height difference

221   between two points of the profile separated by a distance $\Delta x$. For a self-affine profile, the

222   correlation function follows a power-law such that $C(\Delta x) \propto \Delta x^H$ where $H$ is the Hurst

223   exponent.

224   *3.3 Standard deviation of the correlation function (RMS-COR) method*

225   For a profile $L(x)$ containing $N$ points, the height difference $\Delta L$ between each couple of

226   points separated by a distance $\Delta x$ is calculated. The window size $\Delta x$ is varied between the

227   sampling distance and the size of the system and, for a given $\Delta x$, the standard deviation of

228   the height difference $\sigma(\Delta L_{\Delta x})$ is calculated. For a self-affine surface this measurement

229   follows a power-law such that $\sigma(\Delta x) \propto \Delta x^H$. This method was successfully applied to

230   characterize the self-affine properties of the Vuache fault plane (RENARD *et al.* 2006).

231   *3.4 Fourier power spectrum (FPS) method*

232   The Hurst exponent $H$ can be estimated from the Fourier power spectrum which has a power

233   law form for a 1D self-affine profile (BARABASI and STANLEY, 1995; MEAKIN 1998).

234   For each parallel profile, the Fourier power spectrum $P(k)$, *i. e.* the square of the modulus of

235   the Fourier transform, is calculated as a function of the wave-number $k$. Then the spectrum of

236   the whole surface is calculated by stacking all the 1D Fourier transforms to reduce the noise

237   associated with individual profiles. For each profile of length $L$ containing $N$ increments,

238   the spatial frequencies range between $1/L$ and the Nyquist frequency $N/2L$ (*i.e.* the

239   reciprocal of the interval between data points). When plotting the power spectrum as a

240   function of $k$ in log-log space, a self-affine function reveals a linear slope, which is itself a

241   function of the Hurst exponent $H$ through $P(k) \propto k^{-1-2H}$.

242  *3.5 Average wavelet coefficient power spectrum (WPS) method*

243  The average wavelet coefficient method consists of decomposing the input signal into
244  amplitudes that depend on position and scale. The wavelet transform of each 1D profile $L(x)$
245  is defined as $W_{a,b} = \frac{1}{\sqrt{a}} \int_{-\infty}^{+\infty} \psi\left(\frac{x-b}{a}\right) |L(x)| dx$ where $\psi$ is the wavelet function. Then the
246  wavelet coefficients are averaged over the translation factor $b$ for each length scale $a$:
247  $W_a = \langle W_{a,b} \rangle_b$. If the profile is self-affine, the wavelet transform verifies statistically that, for
248  any positive dilatation factor $\lambda$, $W_{a,b}[L(\lambda x)] = \lambda^H W_{a,b}$. Accordingly, the averaged wavelet
249  coefficients scale as $W_a \propto a^{H+1/2}$. A wide range of wavelet functions can be used. For a
250  simple and efficient implementation, we chose the Daubechies wavelet of order 12 as
251  suggested in SIMONSEN *et al*. (1998).

252

253  *4. Quantitative estimation of the accuracy of roughness analysis methods*
254  *4.1 Synthetic isotropic and anisotropic rough surfaces*
255  Figures 1a and 1b display the topography of synthetic rough surfaces where the data set
256  includes 2049 × 2049 points regularly spaced on a grid. Figure 1a shows an isotropic rough
257  surface, whereas Figure 1b shows an anisotropic surface, with corrugations elongated parallel
258  to the direction of smaller Hurst exponent (analogue to the direction of slip on a natural fault
259  surface) and covering a wide range of scales.
260  The roughness amplitude of a profile parallel to the striation direction (green curve in Figure
261  1c) is smaller than that of a perpendicular profile. The profile extracted along the direction
262  with the smallest exponent (green curve) appears more jagged at small scales compared to a
263  perpendicular profile, showing the different effects of the anisotropy of the surface on the
264  waviness and amplitude of the profiles.

265  The outputs of the statistical methods described in section 2.3 are represented on Figure 2
266  Each curve is calculated by averaging the outputs of all possible parallel 1D profiles extracted
267  from the anisotropic surface of Figure 1b. The results are represented in a log-log plot,
268  allowing visualizing the linear portion of the curve that characterizes a power-law distribution
269  (Figure 2). This linear portion is binned in a small number of increments, and a power-law fit
270  is performed to extract the Hurst exponent that characterizes the self-affinity of the profile.
271  The best fits are performed for each curve and a value of the "output" self-affine exponent is
272  then calculated for all the six signal processing methods.
273  Using the RMS correlation function, we have also extracted sets of parallel profiles in several
274  directions, at an angle $\theta$ to the direction of the striations. For each set of profile, we have
275  calculated the correlation function and estimated the value of $H$. The angular dependence of
276  $H$ could be represented on a polar plot (inset in Figure 2d) (RENARD *et al*., 2006). The
277  anisotropy of such plot characterizes the anisotropy of the surface: an isotropic surface is
278  represented as a circle of radius $H$, whereas an anisotropic one has a more complex elliptical
279  shape.

280  *4.2 Isotropic surfaces: effect of size and input exponent on the output estimation of the Hurst*
281  *exponent*

282  The comparison between the input Hurst exponent used to calculate an isotropic synthetic
283  surface and the output Hurst exponent estimated using the six different methods is represented
284  on Figure 3, for different system sizes. The RMS, MM, COR and RMS-COR methods are all
285  mainly sensitive to the value of the input self-affine exponent (the typical trend of the curve is
286  not parallel to the diagonal). Small self-affine exponents are systematically overestimated
287  whereas large exponents are underestimated. In contrast, the error for the WPS method is
288  mainly function of the system size (the response is more or less parallel to the diagonal). The
289  FPS method appears the most accurate technique, with only slight sensitivity to the input self-

290 affine exponent and size effects. This conclusion should however be interpreted cautiously as
291 the algorithm used to generate the synthetic surface is based on a Fourier transform approach.
292 The conclusion of this comparison tests is that the FPS, WPS, and RMS-COR methods are the
293 most reliable because they have a small dependence on the value of the input Hurst exponent
294 and a slight dependence on system size.

295 *4.3 Anisotropic surfaces: Interaction between the two input roughness exponents*

296 For synthetic self-affine anisotropic surfaces, we have calculated the error on the estimation
297 of the two Hurst exponents. For this, we have built surfaces (2049x2049 points, similar to
298 Figure 1b) for which the Hurst exponents $H_{input //}$ and $H_{input \perp}$ in two perpendicular directions
299 were varied in the range [0.4 – 0.9]. We have then used the six signal processing techniques to
300 estimate the values of these same exponents. The absolute error in the estimation of each
301 Hurst exponent (Figure 4) depends on the input value of these parameters and also on the
302 amplitude of their difference ($H_{input //} - H_{input \perp}$).

303 This error is particularly large for the RMS (up to 20%), MM (up to 25%), and COR (up to
304 35%) methods. When the input anisotropy ($H_{input //} - H_{input \perp}$) increases, the absolute error on
305 the two output exponents increases accordingly. The absolute error is smaller in the direction
306 of the smallest exponent (analogue to the direction of striation on a natural fault surface) than
307 perpendicular to it. Moreover, it is also noteworthy to mention that these three techniques
308 show significant errors in the estimation for input exponents close to 0.8-0.9 even if the
309 anisotropy is minimal, demonstrating the limited reliability of these methods to detect an
310 exponent close to one.

311 The RMS-COR analysis is also sensitive to the input anisotropy (Figure 4d), however such an
312 effect is not strongly pronounced (the absolute errors are smaller, up to 15%). For this
313 method, the error does not depend on the values of the two input Hurst exponents. For
314 example, with a synthetic surface defined by $H_{input //} = 0.8$ and $H_{input \perp} = 0.6$, the absolute error

315 in the estimation of each Hurst exponent is almost identical. As shown in Figure 4d, when
316 anisotropy is small, the errors do not increase significantly for input values close to 1 unlike
317 the three previous methods.

318 The FPS and WPS analysis are only slightly sensitive to the "input" anisotropy and the
319 estimated Hurst exponents do not depend on the input exponent values. Our analysis clearly
320 shows that the FPS, the WPS and, to a lesser extent, the RMS-COR methods are the most
321 reliable. More precisely, the RMS-COR and the WPS techniques slightly underestimate and
322 overestimate, respectively, the roughness exponent compared to the FPS method.

323 We have also analyzed the azimuthal dependence of the Hurst exponent for synthetic
324 anisotropic self-affine surfaces. Comparisons of the "output" anisotropy estimated using the
325 RMS-COR method and the "input" anisotropy is represented on Figure 5. We have used this
326 technique because it does not require interpolation of the profiles, whereas the FPS and WPS
327 methods would need regularly spaced data point. A significant directional morphological
328 anisotropy of surfaces is visible on these polar plots of $H$ even if a low "input" anisotropy is
329 imposed, thus demonstrating the reliability of the RMS-COR method to detect a slight
330 morphological anisotropy. Remarkably, following a 360° rotation, the azimuth variation of $H$
331 is not progressive. When departing a few degrees from the direction of the smallest "input"
332 exponent, the "output" exponent is already very close to the largest "input" exponent. This
333 property of anisotropic self-affine surface is not well understood yet.

334 A tentative way to calculate the error on the anisotropy that is made when estimating the
335 anisotropy of the surface $|H_{//} - H_\perp|$ is represented on Figure 6. This plot indicates the error
336 on the estimation of the anisotropy of the surface, and therefore provides some bounds on the
337 accuracy of the determination of this property. Almost all methods underestimate the
338 anisotropy, except the Fourier power spectrum which slightly overestimates it. For the RMS,
339 MM, and COR methods, when the "input" anisotropy $|H_{//} - H_\perp|$ increases, the absolute error on the "output" anisotropy increases accordingly. Moreover, this absolute error is similar for all surfaces with the same "input" anisotropy, whatever the values of the two "input" self-affine exponents.

The determination of the "output" anisotropy with the RMS-COR, FPS, and WPS methods is less sensitive to the "input" anisotropy, except for the highest anisotropy, thus demonstrating the robustness of these three methods to determine the azimuth dependence of the statistical properties of an anisotropic self-affine surface. More precisely, estimates reported for the WPS method are somehow systematically lower than the two others techniques.

*4.4 Effect of noise*

In all physical measurements, noise is present in the data because of the limited resolution of the measuring device. Such noise is usually described using Gaussian statistics with a zero mean and a constant variance. We have analyzed how the presence of noise in synthetic data could alter the estimation of the Hurst exponent. For this, we have calculated synthetic anisotropic surfaces and added a Gaussian noise with a standard deviation equal to 1/200 of the standard deviation of the rough surface (Figure 7a, b). This artificial alteration of the synthetic surface mimics measurements biases obtained on natural data, for example due to electronic noise in the measurement device or due to weathering of the fault surface. We have then estimated the Hurst exponents using the six signal processing methods and compared the results with the noise-free analysis. The results confirm that adding noise to the synthetic data induces a leveling-off of the curves at small length scales (RENARD *et al.*, 2006; SAGY *et al.*, 2007) and therefore a possible underestimation of the Hurst exponent, for all the six signal processing methods (Figure 7c-h).

Despite the fact that the Gaussian white noise added is isotropic, each plot (Figure 7c-h) indicates that the effect of noise is slightly dependent on the azimuth of the profile: the underestimation of the Hurst exponent is more pronounced along striations than

365 perpendicularly to them. Indeed, the addition of noise in the rough signal preferentially alters

366 the roughness at small scales, and therefore has a stronger effect on the profiles parallel to the

367 striations because they are characterized by a smaller amplitude at large length scales

368 compared to the profiles perpendicular to the striations.

369 For the RMS, MM and RMS-COR methods, the noise does affect not only the small length

370 scales but also the large length scales. Indeed, such an effect is strongly pronounced for these

371 three methods and, slopes of the curves in Figure 7c-f lead to a significant underestimation of

372 the actual value of the self-affine exponents. Notably, the polar plot of $H$ from a surface with

373 added noise obtained with the RMS-COR technique (see Figure 7f) shows errors of 10 % and

374 20 % on the Hurst exponent measured in directions perpendicular and parallel to striations,

375 respectively.

376 Conversely, the COR, FPS, and WPS techniques are less sensitive to the addition of noise. At

377 large scale, the noise appears as a negligible correction, and even if the curves are affected at

378 small scales, the estimation of the self-affine exponent is less affected.

379 *4.5 Effect of missing data*

380 When considering natural fault measurements, local weathering and/or the presence of

381 vegetation may form patches of missing data. To study their influence on estimation of fault

382 surface properties, we generated incomplete data sets removing an increasing percentage of

383 clusters of points from a synthetic surface that initially contained 513x513 points (Figure 8).

384 For the FPS and WPS methods, the incomplete cloud of points was interpolated across the

385 gaps (Figure 8b), using a linear fit. However, for the RMS-COR method, the biased data can

386 be used without interpolation of the holes (Figure 8c).

387 Typically, in our natural data sets 5 % of interpolated holes is the maximum percentage of

388 spurious points removed from the raw scanner data. The results (Figure 8d-f) indicate that the

389 RMS-COR, FPS and WPS analysis show an error of only 4 % on the Hurst exponent estimated on a surface with 40 % holes compared to a complete surface. Therefore, 5% of missing points does not affect significantly the measurement of surface properties, whatever the technique employed.

## 5. Acquisition of roughness data on natural faults at various scales

*5.1 Acquisition of the data on the field and in the laboratory*

The roughness data of several fault samples were acquired at various scales using five different scanning devices (Table 1). At the laboratory scale, we used a home-made laser profilometer (MEHEUST, 2002), where a sample, set on a 2-axis moving table, is scanned by measuring the distance between the sample and a laser head. The horizontal scanning steps are either 20 or 24 micrometers and the vertical resolution is better than 1 micrometer.

On field outcrops, we measured several surfaces with four different LIDARs, where two main technologies were used. The S10 system (Table 1) contains a laser source and two cameras; the distance between the laser head and a surface point is measured by triangulation. The maximum shooting distance is around 15 m and the resolution in the distance measurement is close to 0.5 mm. Surfaces of several square meters can be measured with this system. The main drawback of this system is that it must be operated during night time otherwise the day light may blind the cameras.

The other three LIDAR systems (Table 1) are based on the same technology and were built by three different companies: a light pulse is sent from a laser head and the time of flight to the target point is measured, allowing calculating the distance, knowing light velocity. The whole target surface is scanned by rotating the laser head at constant angular velocity. The main advantages of this technology is that fast scanning rates can be achieved (up to 5000 points/s), the shooting distance can be as large as 1500 m and the system can be operated even under day light. However, compared to the S10 system, the measurement accuracy is lower, between 1 and 2 cm. Note also that if the laser wavelength is in the infra-red range, absorption by water present on the target surface might also alter the quality of the data.

We have used these scanning measurement devices on two faults in limestones, where outcrop fault planes were scanned at various scales. Hand samples of slip surfaces were also collected and measured with the laboratory profilometer.

*5.2. The Magnola normal fault*

The Magnola fault outcrop, in the Fuccino area, is part of the extensive fault system in Central Apenines (Italy). This 15-km long normal fault shows microseismic activity and offsets limestone beds with a vertical displacement larger than 500 meters and a slight shear component witnessed by mechanical striations dipping at a 85° angle on the fault plane. This fault is characterized by recent exhumation (PALUMBO *et al.* 2004, CARCAILLET *et al.* 2008) and the last earthquakes have risen to the surface a ~10-m thick band of fresh limestone (Figure 9a) where mechanical striations and grooves at all scales are still visible and less altered by weathering than older exhumed portions of the fault. We have scanned several sub-surfaces of the fault wall (Table 2, Figure 9b) and collected one hand sample from the roof of the fault for laboratory measurements (Table 2, Figure 9c). This sample, that shows mechanical striations, was dig from below the ground surface, to get a slip surface preserved from climatic weathering. The larger outcrop surfaces show evidence of erosion and some karstic water outlets provided holes for vegetation. Small bushes and grass outcrops had to be removed either directly from the fault plane, of from the LIDAR data. The result was incomplete data sets of the fault surface, with missing points in the records. Nevertheless, elongated bumps and depressions at large scales (Figure 9a), and grooves and ridges at small scales (Figure 9b) aligned parallel to slip can be observed.

440  *5.3. The Vuache strike-slip fault*

441  The Vuache fault is an active strike-slip fault system in the western part of the French Alps

442  (THOUVENOT *et al.* 1998). The fault outcrop we analyzed lies on a short segment that

443  connects to the main Vuache fault and that is not active anymore. This fault was analyzed in

444  RENARD *et al*. (2006) and we present here new data of large and small scale slip surfaces

445  (Figure 10).

446  This fault has a mainly strike-slip component, witnessed by large elongated bumps and

447  depressions associated with linear striations of smaller size observed at all scales up to the

448  resolution of the scanners LIDAR (Figure 10c, d). Conversely, the laboratory data show that

449  the surface below the centimeter scale appears more polished and only smooth decimeter

450  ridges persist (Figure 10f).

451  The fault offsets meter-scale beds of limestones and the fault plane was exhumed ten years

452  ago by the activity of a quarry. As a consequence, the LIDAR measurements could be

453  performed on fresh surfaces, where weathering was minimum and no vegetation had

454  developed on the fault plane. For these surfaces, the data recovery was excellent, greater than

455  99.5%. We therefore obtained the topography of the surfaces without holes in the data,

456  making the signal processing results reliable.

457

458  *6. Roughness results and interpretation*

459  *6.1 Magnola Fault*

460  The roughness analysis results of the Magnola fault outcrop and hand sample are shown on

461  Figure 11. On each plot (Figure 11a-d) both Lidar data (upper curve), acquired with the

462  Optech scanner (Table 2), and laboratory profilometer data (lower curve) are represented,

463  showing a scaling behavior over 5.5 orders of magnitude of length scales (50 μm to

464  approximately 20 m). Each curve represents an average over a large set of parallel 1D profiles

465  extracted from the DEM of the fault surface shown on Figure 9. The level of noise for the
466  field LIDAR scanner (Table 1) is estimated as the height of the flat part of the spectrum at
467  small length scales and is indicated by the black arrows (Figure 11). The flattening of the
468  scaling behavior at large scales is related to a finite size effect.

469  The FPS and WPS techniques performed along and perpendicular to the slip direction (Figure
470  11a-d) indicate that the power laws can easily be connected for the field and laboratory data,
471  demonstrating the robustness of the scaling behavior. Moreover, our results highlight a
472  significant directional morphological anisotropy over a wide range of scales: the profiles
473  parallel to the slip direction are rougher than perpendicular ones (POWER *et al.*, 1988;
474  POWER and TULLIS, 1991; LEE and BRUHN, 1996; POWER and DURHAM, 1997;
475  RENARD *et al.*, 2006; SAGY *et al.*, 2007). These two methods estimate a similar Hurst
476  exponent perpendicular to the slip orientation ($H_\perp = 0.8$) across the whole range of scales
477  investigated in this study, a property similar to fresh mode I fracture surfaces (POWER *et al.*,
478  1987; SCHMITTBUHL *et al.*, 1995b, BOUCHAUD, 1997). However, the FPS technique
479  indicates a greater anisotropy, $|H_{//} - H_\perp| = 0.2$, where $H_{//}$ represents the Hurst exponent in
480  the direction of slip, than quantified by the WPS method ($|H_{//} - H_\perp| = 0.1$). Moreover, the
481  WPS method overestimates the self-affine exponent along the slip direction ($H_{//} = 0.7$)
482  compared to the FPS method ($H_{//} = 0.6$). An attempted explanation of these last two
483  differences is given by our parametric study of synthetic rough surfaces: the WPS method
484  slightly overestimates the roughness exponents when measuring couples of Hurst exponents
485  in perpendicular directions with range of values similar to those of natural fault surface (0.6 to
486  0.9). Notably, the exponents accuracy with anisotropic surface of 2049 x 2049 points and for
487  two Hurst exponents in perpendicular direction of 0.8 and 0.6 is numerically estimated as -
488  0.01 and -0.06 for the wavelet method, respectively (Figure 4f). For example, an amount of -
489  0.06 should be added to the measured minimal exponent with the WPS analysis to obtain the actual one. Accordingly, on natural fault surface with two perpendicular exponents of 0.8 and 0.6 calculated by the FPS, the estimated Hurst exponent in direction parallel to slip is systematically lower than with the WPS method (Figures 11-12). Moreover, since the error on the output Hurst exponent is greater in direction of slip than perpendicularly to it, consequently the output anisotropy decreases, as observed on natural fault surface (Figure 11-12).

Our estimations obtained on the hand sample with the RMS-COR function (see Figure 11e) show that the minimum Hurst exponent ($H = 0.6$) is at 85°, in the direction of slip, and the maximum Hurst exponent ($H = 0.8$) is almost in the perpendicular direction. These two extremes values of the self-affine exponent correspond also to those determined at all scales by the FPS and WPS methods. At the field scale (see Figure 11e), the minimum ($H = 0.4$) and maximum ($H = 0.7$) Hurst exponents are observed in directions similar to that for the hand sample. However, the results obtained with the RMS-COR method suggest that the roughness exponent of the Magnola fault surface is smaller at the field scale compared to the laboratory scale, regardless of the azimuth. We ascribe this variation to natural weathering (pitting) of the exposed fault surface at short wavelengths, as POWER and TULLIS (1991). Indeed, the section of the Magnola fault surface (Figure 9a) shows an increase of the roughness amplitude at short wavelengths created by weathering. Other sections of the Magnola fault surface, that are not presented in this study, display evidence of strong alteration at short length scales leading to a significant reduction of the Hurst exponent, regardless of azimuth. Conversely, the clean hand sample (Figure 9b), that shows mechanical striations, should represent the actual topography of the fault surface at short length scales, related to the faulting process before the action of climatic weathering. This hypothesis is supported by the fact that the fresh slip plane, scanned in laboratory, displays the same self-affine RMS-COR regimes in the directions parallel and perpendicular to slip than those

515  estimated using the FPS and WPS methods. The increase of roughness at short wavelengths
516  on the field surface due to the erosion appears to be similar to the effect of an additional noise
517  tested on "ideal" synthetic surfaces (Figure 7). In both cases, the Hurst exponent decreases in
518  all directions. Our statistical study led on synthetic surfaces shows that the noise effect is
519  more pronounced when using the RMS-COR technique. To summarize, the roughness scaling
520  estimated using the RMS-COR calculated on the weathered field surface exhibits a decrease
521  of the Hurst exponent in all directions, which is not observed with the FPS and WPS
522  techniques.
523  On the polar plot of $H$ obtained at the laboratory scale (see Figure 11e), when departing a few
524  degrees from the direction of slip, the Hurst exponent is close to the value of the maximum
525  Hurst exponent measured in the direction perpendicular to mechanical striations. Such
526  behavior is also visible on 'ideal" synthetic surfaces (Figure 5). In other words, the azimuth of
527  the maximum Hurst exponent is not well-defined (gray shadows on Figure 11e), while the
528  minimum exponent corresponds to a specific orientation. Note that this property is less visible
529  on the polar plot of the altered field fault section (see Figure 11e) where the angular variation
530  of $H$ is more progressive.
531  *6.2 Vuache Fault*
532  The FPS and WPS techniques highlight a significant directional morphological anisotropy
533  over six orders of spatial scales (Figure 12a-d). Profiles parallel to slip have a smaller Hurst
534  exponent than perpendicular ones (POWER *et al.*, 1988; POWER and TULLIS, 1991; LEE
535  and BRUHN, 1996; POWER and DURHAM, 1997; RENARD *et al*., 2006; SAGY *et al*.,
536  2007). These two methods, applied to series of profiles perpendicular to the direction of slip,
537  indicate that the power laws of individual surfaces can easily be connected across the wide
538  range of scales investigated (Figure 12b, d), and the value of $H_\perp = 0.8$ is similar to what was
539  measured on the Magnola normal fault surface. However, in the slip parallel direction there is

540  a slight change in the magnitude of $H_{//}$ located in the length scale range between 5 mm and 2
541  cm (gray shadows on Figure 12a, c). $H_{//} = 0.75$ is larger below this length scale range than at
542  larger length scales, where $H_{//} = 0.65$ and $H_{//} = 0.7$ for the FPS and WPS methods,
543  respectively. The smoother aspect of the Vuache fault surface in the direction of slip
544  compared to the perpendicular direction is an obvious and expected consequence of striations,
545  but the smallest directional morphological anisotropy at the laboratory scale compared to the
546  field scale is novel in this study. At the field scale, the morphology of the elongated bumps
547  and depressions along the slip direction is different from the grooves and striations observed
548  at the laboratory scale (Figure 10). As the fault Vuache outcrop is quite fresh, and was
549  preserved from the climatic erosion, we propose that this cross-over in the slope at length
550  scale of several millimeters is significant. We interpret this cross-over length scale as a
551  witness of different mechanical processes responsible for the creation of fault topography at
552  different spatial scales.

553  Our results obtained at the field and intermediate scales with the RMS-COR function (see left
554  and center plots on Figure 12e) show that the minimum Hurst exponent ($H_{//} \in [0.55 - 0.6]$) is
555  oriented at 20° with respect to the horizontal, indicating the fault had a normal component and
556  not only a strike-slip one. The maximum value $H = 0.75$ is located for an almost
557  perpendicular direction. These two extremes values of the self-affine exponent are slightly
558  lower when estimated using the RMS-COR function than when using the FPS or the WPS
559  methods. This slight underestimation with the RMS-COR technique is consistent with our
560  results on synthetic surfaces for the accuracy in this range of parameters (Figure 4). Indeed,
561  our parametric study led on anisotropic synthetic surfaces shows that the estimation of $H$
562  calculated with the RMS-COR technique slightly underestimates its actual value (Figure 4d).
563  At the laboratory scale (see right plot on Figure 12e), $H_{//} = 0.8$ and $H_{\perp} = 0.9$ are located in
564  orientations similar to that for the three larger surfaces measured on the field. However, the

565  polar plot of $H$ calculated using the RMS-COR function suggests an increase of the scaling
566  exponents in all directions, while the estimation using the FPS or WPS techniques underlined
567  that the exponent increased only along the slip direction. In addition, the two exponents $H_{//}$
568  and $H_\perp$ for the hand sample are overestimated when calculated with the RMS-COR method
569  compared to the FPS and WPS techniques. The latter observation cannot be explained by the
570  results of our parametric study on synthetic surfaces. As a consequence of the extremely
571  smooth aspect of the hand sample at small scales (Figure 10f), the RMS-COR method could
572  lose its robustness. However, a new reliable result is that the directional morphological
573  anisotropy calculated by the RMS-COR function significantly decreases at the laboratory
574  scale, as observed with the FPS and WPS techniques.

575  As already observed on the polar plot of $H$ calculated on the Magnola fault surface (Figure
576  11e), the azimuth of the maximum Hurst exponent is less constrained (gray shadows on
577  Figure 12e) than the single specific orientation of the striations. Remarkably, despite the weak
578  anisotropy of the hand sample topography, the slip direction is always significant,
579  demonstrating the accuracy and reliability of the RMS-COR method.

580

581  *7. Discussion & Conclusion*

582  The six statistical tools used in this study have different response under the effect of two kinds
583  of biases, the intrinsic errors of the methods (Figures 3, 4, 5, 6) and the artifacts inherent in
584  data acquisition (Figures 7, 8). Using a parametric approach, where we varied the size of the
585  surface and its anisotropy, we selected the three most reliable and accurate methods (RMS-
586  COR, FPS, WPS) for roughness analysis of natural fault topography (Figures 11, 12). The
587  Hurst exponent estimation at various scales for the natural slip surfaces displays the same
588  trends and provides a consistent and robust characterization of their scaling regimes. We emphasize that the slight variations in the results given by each one of these methods fall within the range of error estimated by our parametric study (see Section 4).

One of the most robust results of our scaling analysis is that the FPS and the WPS methods estimate a same Hurst exponent equal to 0.8 in the direction perpendicular to slip, over approximately 6 orders of magnitude of length scales for two different fault surfaces (Figures 11b, d and 12b, d). However, in the slip direction two different scaling behaviors are highlighted: the Magnola fault surface shows an identical scaling regime at large and small scales (Figure 11a, c). Conversely, the scaling property of the Vuache fault roughness exhibits a cross-over in the slope at length scale of several millimeters (Figure 12a, c). In other words, the scaling property of this fault surfaces is similar at large scales but changes at small scales. The statistical analyses (Figure 12) and the scan of the Vuache fault surface (Figure 10) clearly show a smoothing of the roughness below a length scale of several millimeters.

The scaling regime of 0.8 measured in the direction perpendicular to slip is a classical result already observed for tensile cracks (POWER *et al*., 1987; SCHMITTBUHL *et al*., 1995b, BOUCHAUD, 1997), indicating that the topography of the fault surface in the direction perpendicular to slip has not registered the effect of shear. Along slip, the general interpretation is that mechanical wear processes, such as frictional ploughing, cause striations that reduce the amplitude of the large scale roughness (POWER *et al.*, 1987, 1988; POWER and TULLIS, 1991; POWER and DURHAM, 1997; SAGY *et al.,* 2007) and accordingly the Hurst exponent. Nevertheless, our scaling analysis seems to indicate different mechanical processes responsible for the creation of fault topography at different spatial scales.

Prior comparative studies of natural fault roughness based on 1D profilometry (POWER and TULLIS, 1991; LEE and BRUHN, 1996) suggest a change in scaling properties between large and short length scales. However, due to technical limitations, their measures were not sufficiently accurate to decipher if this variation was related either to small-scale surface weathering of the fault scarp or to the faulting process itself.

From laboratory experiments CHEN and SPETZLER (1993) suggest that the break in slope at length scales of several millimeters is caused by a change in the dominant mode of deformation from small-scale intergranular cracking to intragranular cracking at a larger scale. We think this interpretation does not apply to the Magnola and Vuache faults because the grain scale of these limestones is very small (< 0.1mm), well below the observed cross-over length scale.

Recently, SAGY *et al.* (2007) observed that faults with large cumulated slip display surfaces with elongate, quasi elliptical bumps at field scale and are polished at small scales. Conversely, fault with a small cumulative slip are rough on all scales. SAGY and BRODSKY (2008) proposed that the waviness of the large slip fault surface reflects variations of thickness of the cohesive layer under the slip surface formed as boudinage structures (JOHNSON and FLETCHER, 1994; SMITH, 1977; TWISS and MOORES, 1992, GOSCOMBE *et al.*, 2004). Therefore, they evoke two different deformation mechanisms between large and small scales: abrasion caused by frictional sliding at the origin of the smoothing at small scales, and "boudinage" creating elongated bumps and depression at large scales.

From our study of natural fault roughness, we observe large elongated bumps and depressions in slip direction on two different fault planes (Figures 9b, 10a-d). There is no evidence that the small segment, polished at small scale (Figure 10f), that connects to the main Vuache fault has accumulated more slip than the Magnola fault surface that is rough at all scales (Figure 9b, c). Therefore, we propose that large elongated bumps and depressions of several meters in length with maximum amplitude of around 2 m may reflect the processes of lateral growth and branching that links together several fault surfaces, during all the stages of the evolution

638  of the fault zone, as suggested by LEE and BRUHN (1996) and LIBICKI and BEN-ZION
639  (2005).
640  At small scales, two different kinds of scaling regime are observed on the two fault surfaces,
641  (Figures 11, 12), both being linked to mechanical wear process. Frictional sliding is expressed
642  through ploughing of small asperities and is responsible for the small scale abrasional
643  striations on the Magnola fault surface (Figure 9c). This mechanism is also responsible for the
644  polishing of the Vuache fault surface below the centimeter scale (Figure 10f). One should
645  keep in mind that only one hand sample was measured for each surface and therefore it is
646  possible that differences of the scaling behavior between the two fault planes reflect spatial
647  heterogeneity of core fault at small scales. Notably, on the Vuache fault, although the surface
648  appears polished at the laboratory scale on the whole surface, zones with striations due to
649  ploughing elements could be present. A more extensive study of fault roughness in several
650  different faults should therefore bring more information on the mechanisms at work during
651  faulting.
652
653
654
655
656
657
658
659
660
661
662

Table 1: Characteristics of the field and laboratory scanner devices.

| **3D scanner device** | S10 | GS100 | LMS Z420i | Ilris-3D | Lab. Profilometer |
|---|---|---|---|---|---|
| **Company** | Trimble | Trimble | Riegl | Optech | Univ. Strasbourg |
| **Resolution (dx)** | 0.5 mm | 10 or 20 mm | 20 mm | 20 mm | 20 or 24 $\mu$m |
| **Noise on the data** | 0.9 mm | 4.5 mm | 10.2 mm | 20 mm | < 1 $\mu$m |
| **Acquisition speed** | 70 pts/s | 5000 pts/s | 5000 pts/s | 2500 pts/s | 60 pts/s |

Table 2: Fault surfaces analyzed in this study.

| Fault | **Surface area, dx** | **Scanner** |
|---|---|---|
| Vuache, SURF1 | 17 x 10 m, 20 mm | GS100 |
| Vuache, SURF7 | 24 x 11 m, 30 mm | GS100 |
| Vuache, SURF6 | 45 x 9 m, 20 mm | LMS Z420i |
| Vuache, SMALL | 10 x 9 cm, 24 $\mu$m | Lab. Profilometer |
| Vuache, SURF-JPG | 0.5 x 0.5 m, 1 mm | S10 |
| Magnola, A32 | 35 x 15 m, 20 mm | Optech |
| Magnola, M2 | 7.2 x 4.5 cm, 20 $\mu$m | Lab. Profilometer |

```matlab
function RoughSurf = Synthetic2DFault(N,H1,H2)

% This Matlab(c) function creates a self-affine 2D surface,
% with a directional anisotropy (courtesy of Hermine Bierme,
% Univ. Paris V, France), when H1 is different from H2.
% Input parameters:
% N = size of the surface: (2^(N+1)+1) x (2^(N+1=+1)
% Typically N must be between 8 and 11 when running on a
desktop computer.
% H1, H2: Hurst exponents in two perpendicular directions
% must be positive, smaller than 1.
% Output result:
% RoughSurf: rough surface of size (2^(N+1)+1) x (2^(N+1=+1),
% with two perpendicular directions of self-affinity
% characterized by Hurst exponents H1 and H2, l1 = 1/H1 and
% l2 = 1/H2 represent the eigen values of the anisotropy
% diagonal matrix
l1 = 1/H1;
l2 = 1/H2;
X = (-2*2^N:2:2*2^N)/(2^(N+1));
X(2^N+1) = 1/2^N;
Y = (-2*2^N:2:2*2^N)/(2^(N+1));
Y(2^N+1) = 1/2^N;
XX = X(ones(1,2*2^N+1),:);
YY = Y(ones(1,2*2^N+1),:)';
clear X Y

% rho is the pseudo-norm associated to the eigen values l1 and
l2
% rho(x,y)=(abs(x)^(2/l1) + abs(y)^(2/l2) )^(1/2)
rho = sqrt(abs(XX).^(2/l1)+abs(YY).^(2/l2));
clear XX YY

%phi is the spectral density of the field built from rho
phi = rho.^(1 + (l1+l2)/2);
clear rho

% W = Fourier transform of the anisotropic Gaussian field
Z = randn(2*2^N+1,2*2^N+1);
W = fftshift(fft2(Z))./phi;
clear Z
T = real(ifft2(ifftshift(W)));
RoughSurf = T-T(2^N+1,2^N+1);

% Plot the 2D rough surface
imagesc(RoughSurf);
axis equal
axis tight
```

871

872    **Figure 1.** Digital Elevation Models (DEM) of 2D synthetic self-affine surfaces (up) and 1D

873    profiles (down) generated using the algorithm of the appendix. (a) Surface with an isotropic

874    self-affine property characterized by a Hurst exponent of 0.8. (b) Anisotropic self-affine

875    surface with two Hurst exponents ($H_{//} = 0.6$ and $H_{\perp} = 0.8$) in perpendicular directions. (c)

876    Representative 1D profiles extracted in two perpendicular directions of surface (b). Inset:

877    magnified portion of a profile along the $H_{//}$ direction (parallel to the striations), which has a

878    statistically similar appearance to the entire profile when using a rescaling $\delta x \rightarrow \lambda \delta x$, $\delta z \rightarrow$

879    $\lambda^H \delta z$.

880

881    **Figure 2.** Outputs of the six signal processing techniques applied on the data of the

882    anisotropic self-affine surface shown in Figure 1b. (a) Root-mean-square correlation (RMS),

883    (b) maximum-minimum height difference (MM), (c) correlation function (COR), (d) RMS

884    correlation function (RMS-COR), (e) Fourier power spectrum (FPS), (f) Wavelet power

885    spectrum (WPS). The inset in (d) displays a polar plot of $H$ obtained by the RMS-COR

886    method and allowing then to determine the azimuth dependence of $H$. The black points

887    correspond to the Hurst exponents for the two profiles shown on this plot.

888

889    **Figure 3.** Comparisons between the "input" Hurst exponents introduced in the construction of

890    isotropic fractional Brownian surfaces and the "output" exponent measured on these surfaces

891    using the six independent signal processing techniques. The effect of system size is also

892    studied. (a) Root-mean-square correlation (RMS), (b) maximum-minimum height difference

893    (MM), (c) correlation function (COR), (d) RMS correlation function (RMS-COR), (e) Fourier

894   power spectrum (FPS), (f) Wavelet power spectrum (WPS). The gray line in (a), (b), and (c)
895   indicate for which input exponent the error in the estimation is closest to zero.

896

897   **Figure 4.** Intrinsic errors in the estimation of the Hurst exponents for anisotropic synthetic
898   surfaces characterized by two Hurst exponents $H_{input //}$ and $H_{input \perp}$ in perpendicular directions
899   (2049x2049 points, similar to Figure 1b). For each signal processing method, histogram plots
900   are represented where the horizontal axis contains the Hurst exponents $H_{input //}$ and $H_{input \perp}$
901   used as inputs to generate the synthetic surface. The vertical axis represents the difference
902   between the input exponent and the estimated output Hurst exponent, using the six different
903   methods. For each method, two histogram plots are represented: the upper one shows the
904   difference ($H_{input //} - H_{output //}$) and the lower one the difference ($H_{input \perp} - H_{output \perp}$). The black
905   top surfaces on the histogram bars indicate a negative difference (overestimation of the output
906   exponent) and the color ones a positive difference (underestimation of the output exponent).
907   (a) Root-mean-square correlation (RMS), (b) maximum-minimum height difference (MM),
908   (c) correlation function (COR), (d) RMS correlation function (RMS-COR), (e) Fourier power
909   spectrum (FPS), (f) Wavelet power spectrum (WPS).

910

911   **Figure 5.** Polar plots of the angular dependence of the two Hurst exponents $H_{//}$ and $H_{\perp}$ for
912   synthetic anisotropic surfaces with principal directions oriented at angles $\theta_{//} = 0°$ and
913   $\theta_{\perp} = 90°$. The Hurst exponents $H_{//,\perp}(\theta)$, as defined by the slope $\beta = H$ in Figure 2d, were
914   calculated on series of 1D profiles extracted at an angle $\theta$ on 2049x2049 points surfaces.
915   Three series of simulations are represented here for three values of $H_{//}$ in the range [0.7 –
916   0.9]. For each polar plot, the different curves corresponds to successive values of $H_{\perp}$. The

917  dashed lines correspond to the values of the output Hurst exponent measured with the RMS
918  correlation method (center dashed circle: $H = 0.2$, outer dashed circle: $H = 1$).

919

920  **Figure 6.** Quantification of the intrinsic errors on the estimation of the anisotropy of the
921  surface ($H_{//} - H_{\perp}$). The difference between the "input" anisotropy (difference between the
922  two "input" Hurst exponents) introduced in synthetic surfaces (2049x2049 points, similar to
923  Figure 1b), minus the "output" anisotropy is represented for the six signal processing
924  methods: (a) Root-mean-square correlation (RMS), (b) maximum-minimum height difference
925  (MM), (c) correlation function (COR), (d) RMS correlation function (RMS-COR), (e) Fourier
926  power spectrum (FPS), (f) Wavelet power spectrum (WPS). Bars with black top surface
927  indicate an overestimation ("input" anisotropy - "ouput" anisotropy < 0) and colored top
928  surfaces indicate an underestimation ("input" anisotropy - "ouput" anisotropy > 0). Gray bars
929  indicate isotropic surfaces ($H_{//} = H_{\perp}$), thus without errors.

930

931  **Figure 7.** Influence of an additional noise on the self-affinity property of an anisotropic
932  synthetic surface (513x513 points) characterized by two Hurst exponents ($H_{//} = 0.6$ and
933  $H_{\perp} = 0.8$) in perpendicular directions. The Gaussian white noise is characterized by a
934  standard deviation two hundred times smaller than the roughness amplitude of the synthetic
935  surface. (a) Example of 1D profiles extracted in two perpendicular directions of an "ideal"
936  surface. (b) The same profiles but with an additional noise. Note that altered profiles appear
937  more jagged or rougher at small scale compared the noise-free profiles. Analyses of those
938  altered surfaces are performed with the six independent self-affine methods: (c) Root-mean-
939  square correlation (RMS), (d) maximum-minimum height difference (MM), (e) correlation
940  function (COR), (f) RMS correlation function (RMS-COR), (g) Fourier power spectrum
941  (FPS), (h) Wavelet power spectrum (WPS). For each method, except the RMS-COR function,

942  two plots are represented: the upper plot shows the difference between the noise-free surface
943  and the altered surfaces in the direction of the largest exponent and the lower one in direction
944  of the smallest exponent. For the RMS-COR technique, the upper plot represents the
945  difference between the noise-free and the altered surface in both directions: perpendicular and
946  parallel to striations. The lower subplot displays two polar plot of $H$ obtained with the noise-
947  free and the biased surfaces. Note that the flattening of the scaling behavior at large scales is
948  related to a finite size effect.

949

950  **Figure 8.** Influence of the interpolation of missing data on the estimation of the Hurst
951  exponent. (a) Synthetic surface (513x513points) with H = 0.8. (b) Same surface with 5% of
952  holes (white dashed lines) that have been interpolated. c) 3D view of the surface in a) with the
953  holes. (d) RMS correlation function (RMS-COR), (e) Fourier power spectrum (FPS), (f)
954  Wavelet power spectrum (WPS). The different curves on each plot present the result of the
955  analysis for five different percentages of missing points (5%, 10%, 20%, 40%).

956

957  **Figure 9.** 3D scanner data of the Magnola fault slip surface at different scales. (a) Photograph
958  of the fault surface, showing significant weathering and covering by vegetation. The black
959  polygon corresponds to the surface shown below. (b) Digital Elevation Model (DEM) of the
960  surface A32 (Table 2). The LIDAR data contain 799,830 points, sampled on a roughly regular
961  grid of ~40 x ~40 mm. The measurements were performed on a ~20 x ~20 mm grid and then
962  averaged on a coarser grid. The resolution of the elevation is 20 mm. The fault surface shows
963  elongated bumps (red) and depressions (blue), dipping at approximately 85°, and indicating a
964  main normal slip motion. The corrugation, with maximum amplitude of around 2.2 m, can be
965  observed at all scales down to the measurement resolution. (c) DEM of hand sample M2
966  (Table 2) that contains 3999 x 3120 points on a regular mesh of 20 x 20 μm$^2$. The surface

967  contains grooves (blue) and ridges (red) aligned parallel to slip and with maximum amplitude
968  of around 1.2 mm.

969

970  **Figure 10.** 3D data of the Vuache fault slip surface at all scales. (a) Photograph of the
971  outcrop, where the white boxes correspond to the surfaces shown in (c) and (d). (b) Lateral
972  photograph of the slip plane that highlights its remarkable waviness. (c-d) Fault surface
973  topography of SURF1 and SURF6 (Table 2). Each surface contains approximately 450,000
974  points sampled on a constant grid of 20 x 20 mm. The resolution of the elevation is 10.2 mm
975  for (c) and 4.5 mm for (d), respectively. The surfaces show large elongated bumps (red) and
976  depressions (blue) with maximum amplitude of around 2 m associated with linear striations of
977  smaller size (grooves and ridges). Both geometrical patterns dipping at 15-25° indicate a
978  strike-slip motion with a small normal component. (e) DEM of the bumpy zone SURF-JPG
979  (Table 2) that contains 107,606 points sampled on a regular grid of 1 x 1 mm. Note the
980  vertical exaggeration. The resolution of the elevation is 0.9 mm. (f) DEM of the hand sample
981  SMALL (Table 2) that contains 4099 x 3333 points on constant grid of 24 x 24 $\mu m^2$. The
982  resolution of the elevation is less than one micrometer. The scans clearly show a smoothing of
983  the roughness from large to small scales. Large fault surface measured on the field have
984  asperities over the entire range of observed scales down to the measurement resolution.
985  Conversely, the laboratory data show that the surface below the centimeter scale appears more
986  polished and only smooth decimeter ridges persist.

987

988  **Figure 11.** Roughness scaling analysis from the best preserved outcrops of the Magnola fault
989  plane, scanned using ground based LIDAR (surface A32, Figure 9b), or using the laboratory
990  profilometer (hand sample M2, Figure 9c). (a-b) Fourier Power Spectrum (FPS), and (c-d)
991  Wavelet power spectrum (WPS) along two directions, parallel and perpendicular to the

992 direction of slip, are represented in log-log plots. Power-law fits (black dotted line) are
993 performed for each curve and the corresponding slopes ($\beta$) and roughness exponents ($H$) are
994 indicated above the spectra. The inset displays an example of the amplitude ($Z$) and the
995 wavelength ($\lambda$) of a rough profile. Contours (red lines) of constant amplitude ($Z$) to
996 wavelength ($\lambda$) ratio, reflecting a self-similar behavior, are provided to allow easier
997 interpretation of the spectra. Black arrows indicate the level of noise of the LIDAR. (e)
998 Surface anisotropy revealed by the angular variation of the Hurst exponent determined by the
999 RMS correlation function method. The polar plot of H on the left and the right sides
1000 correspond to data of the field surface and hand sample shown on Figure 9, respectively.

**Figure 12.** Roughness scaling analysis from five surfaces (Figure 10, Table 2) of the Vuache fault, covering 6 orders of magnitude of frequencies or wavelengths (40 μm to approximately 40 m). The data collected contain four surfaces (blue and green curves) that have been scanned using LIDAR and one hand sample (magenta curve) measured by laboratory profilometer. (a-b) Fourier power spectrum (FPS) and (c-d) Wavelet power spectrum (WPS) along two directions, parallel and perpendicular to the direction of slip, are represented in log-log plots. Power law fits (black dotted line) are shown for each curve and the corresponding slopes ($\beta$) and roughness exponents ($H$) are indicated next to the spectra. Each curve is an average over a series of parallel profiles extracted from the DEM shown on Figure 10. The level of noise for the Lidar scanners is estimated as the height of the flat part of the spectrum at small length scales and is indicated by the black arrows. The flattening of the scaling behavior at large scales is related to a finite size effect. Contours (red lines) of constant amplitude ($Z$) to wavelength ($\lambda$) ratio, reflecting a self similar surface, are also indicated a guide for the eye. The inset displays an example of the amplitude ($Z$) and the wavelength ($\lambda$) of a rough profile. (e) Roughness anisotropy revealed by the angular variation of the

1017   Hurst exponent determined by the RMS-COR method. The polar plots of $H$ on the left, in the
1018   center and on the right correspond to data of the field surface, the intermediate scale section
1019   and hand sample, respectively (see Figure 10).
1020

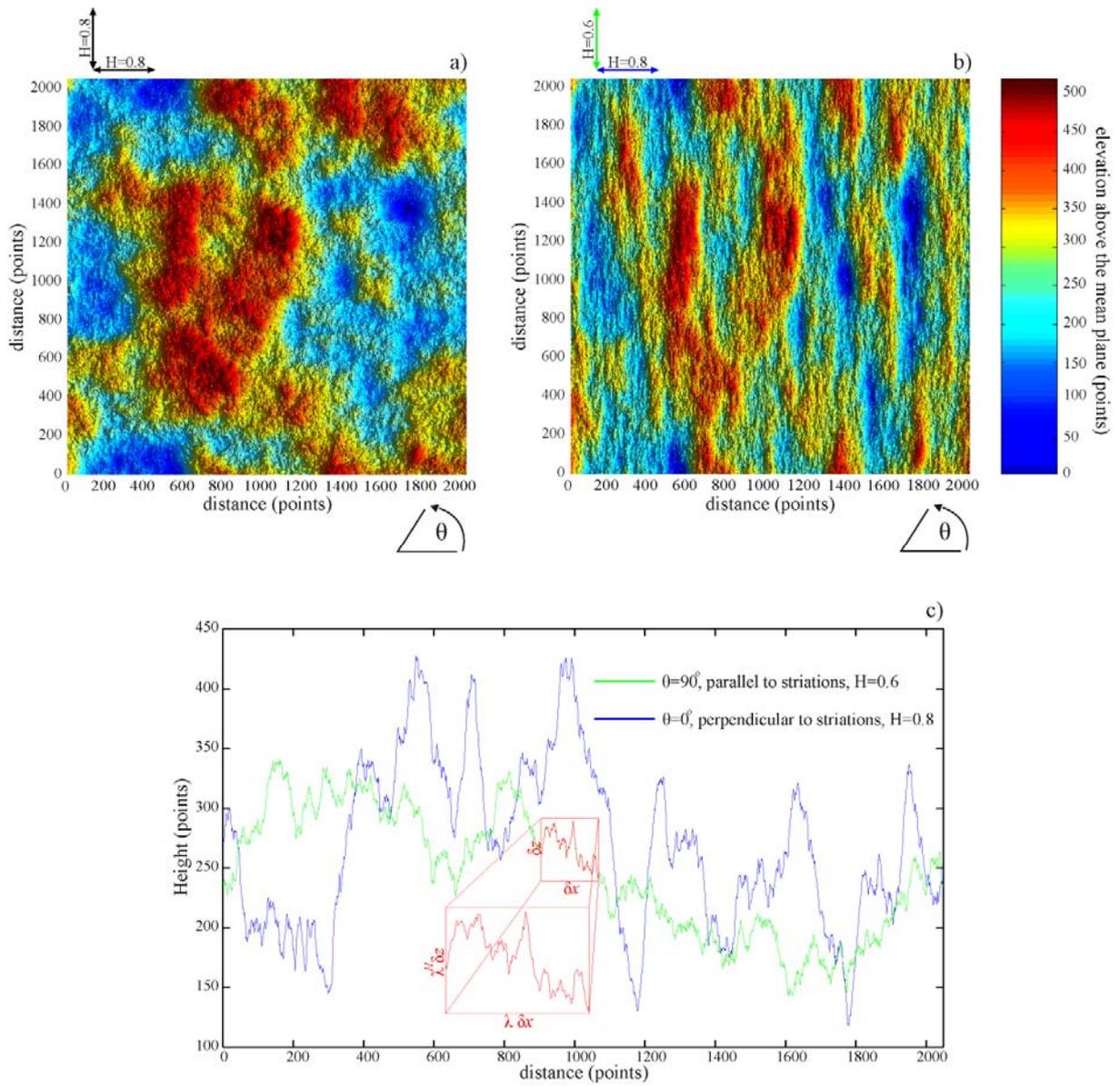

1021

1022 **Figure 1.**

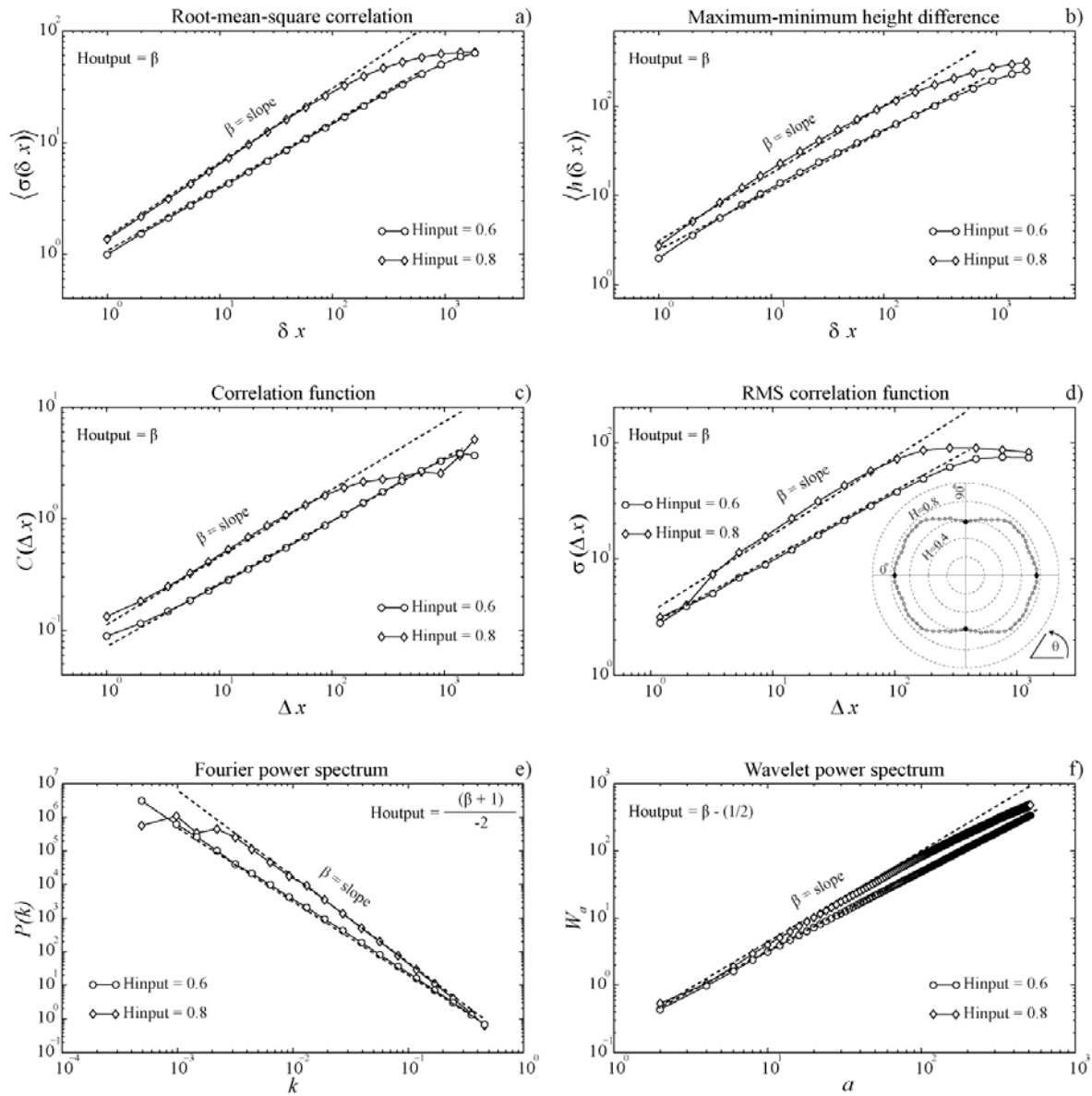

**Figure 2.**

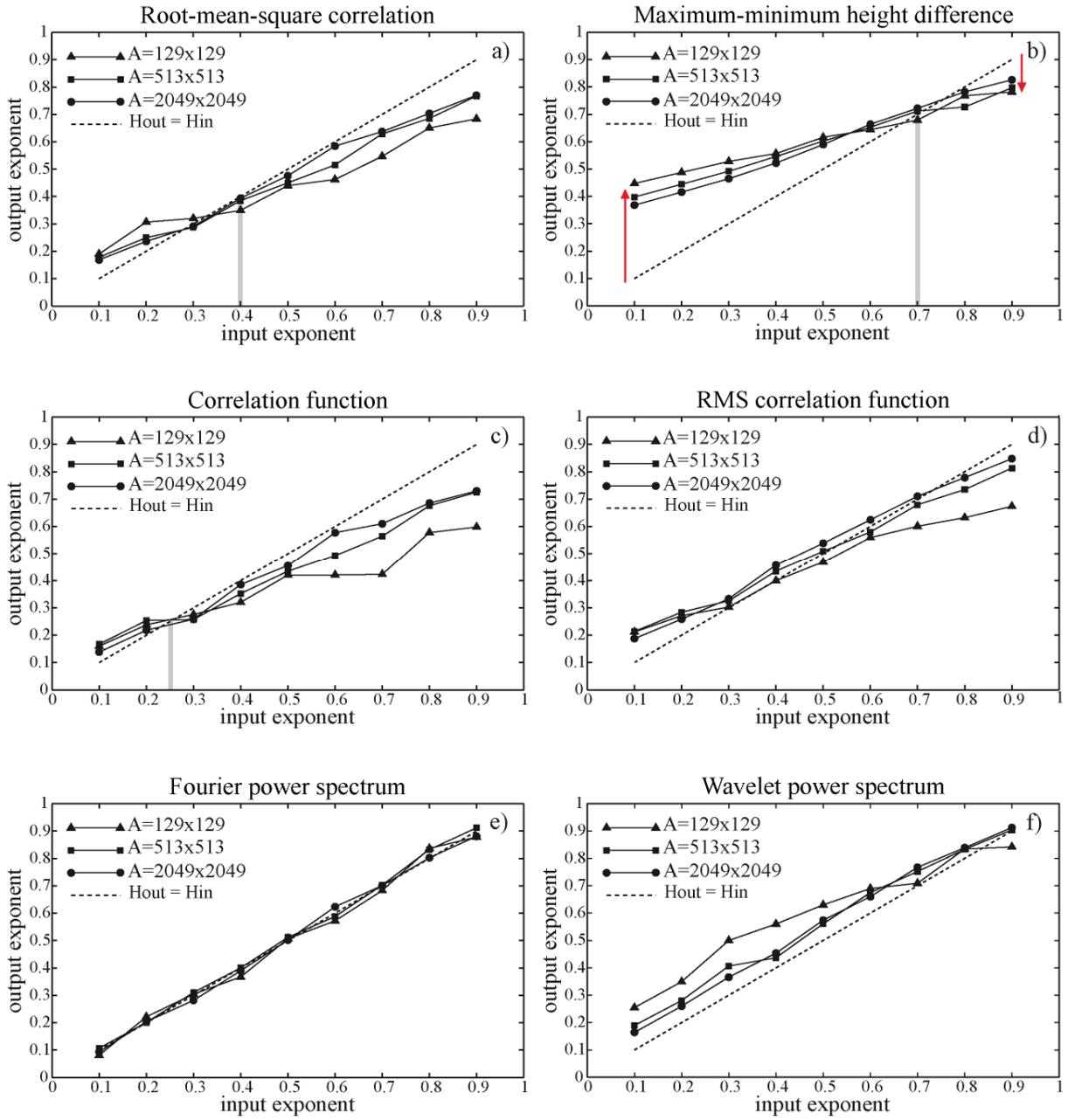

**Figure 3.**

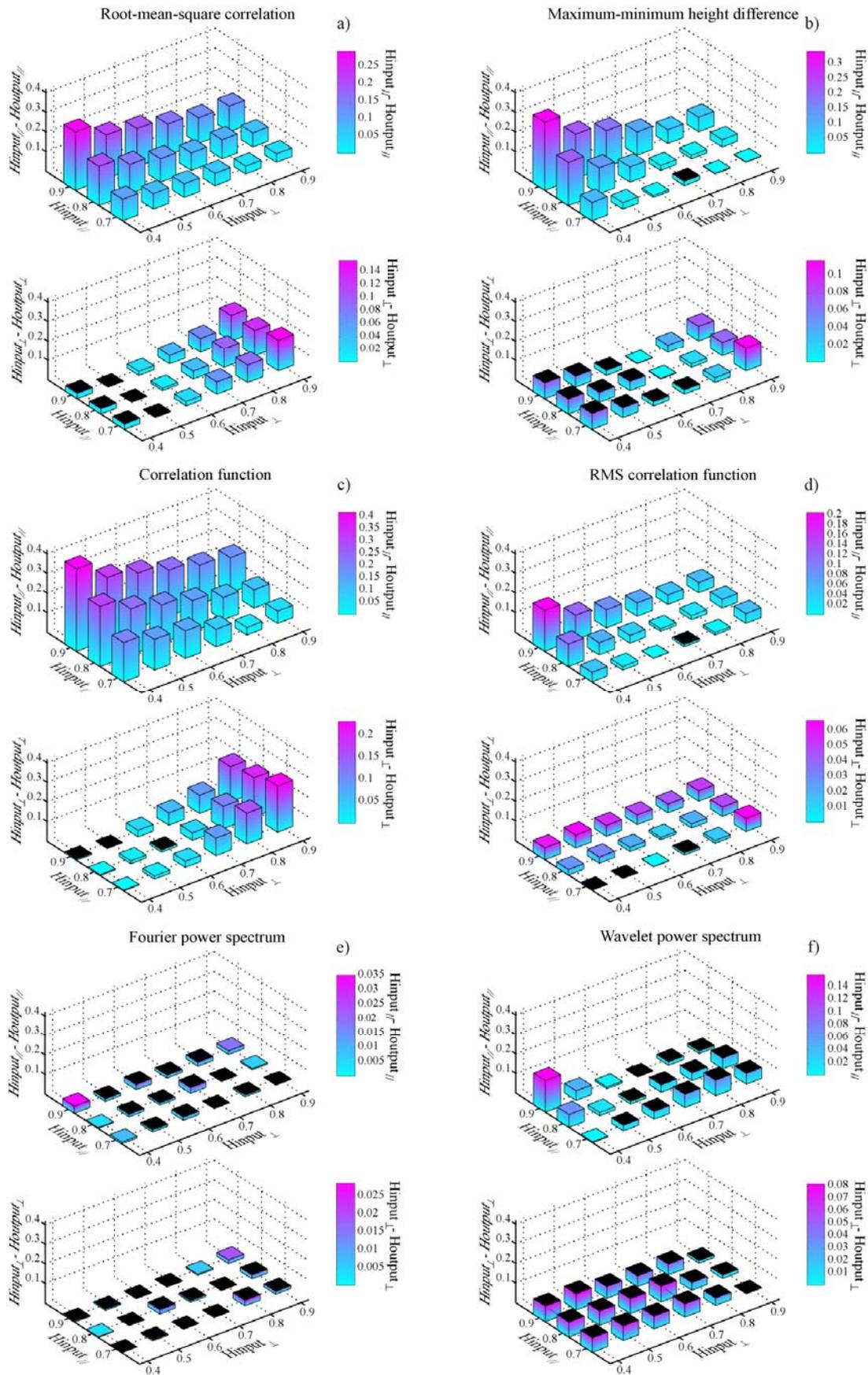

**Figure 4.**

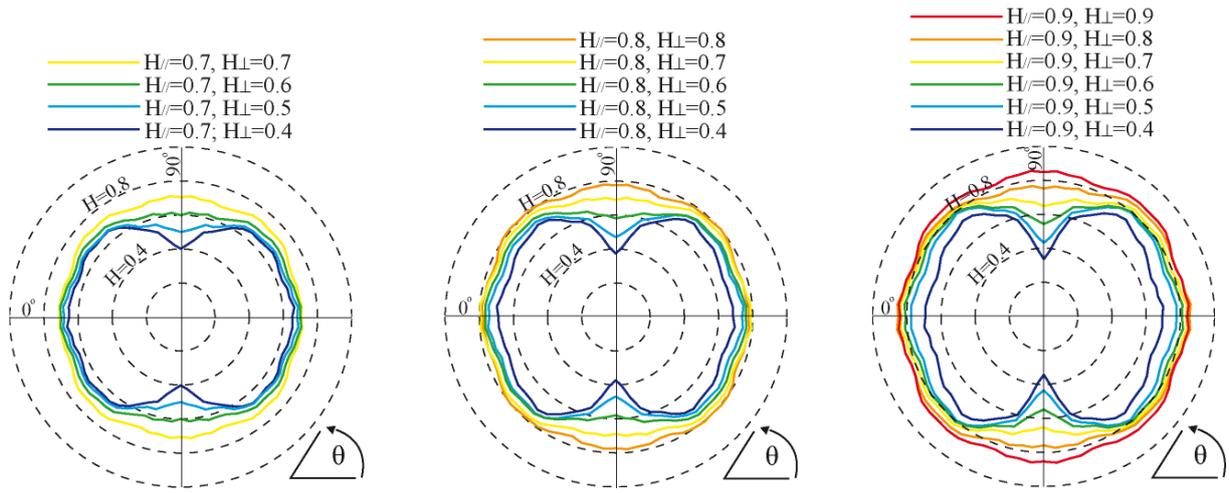

Figure 5.

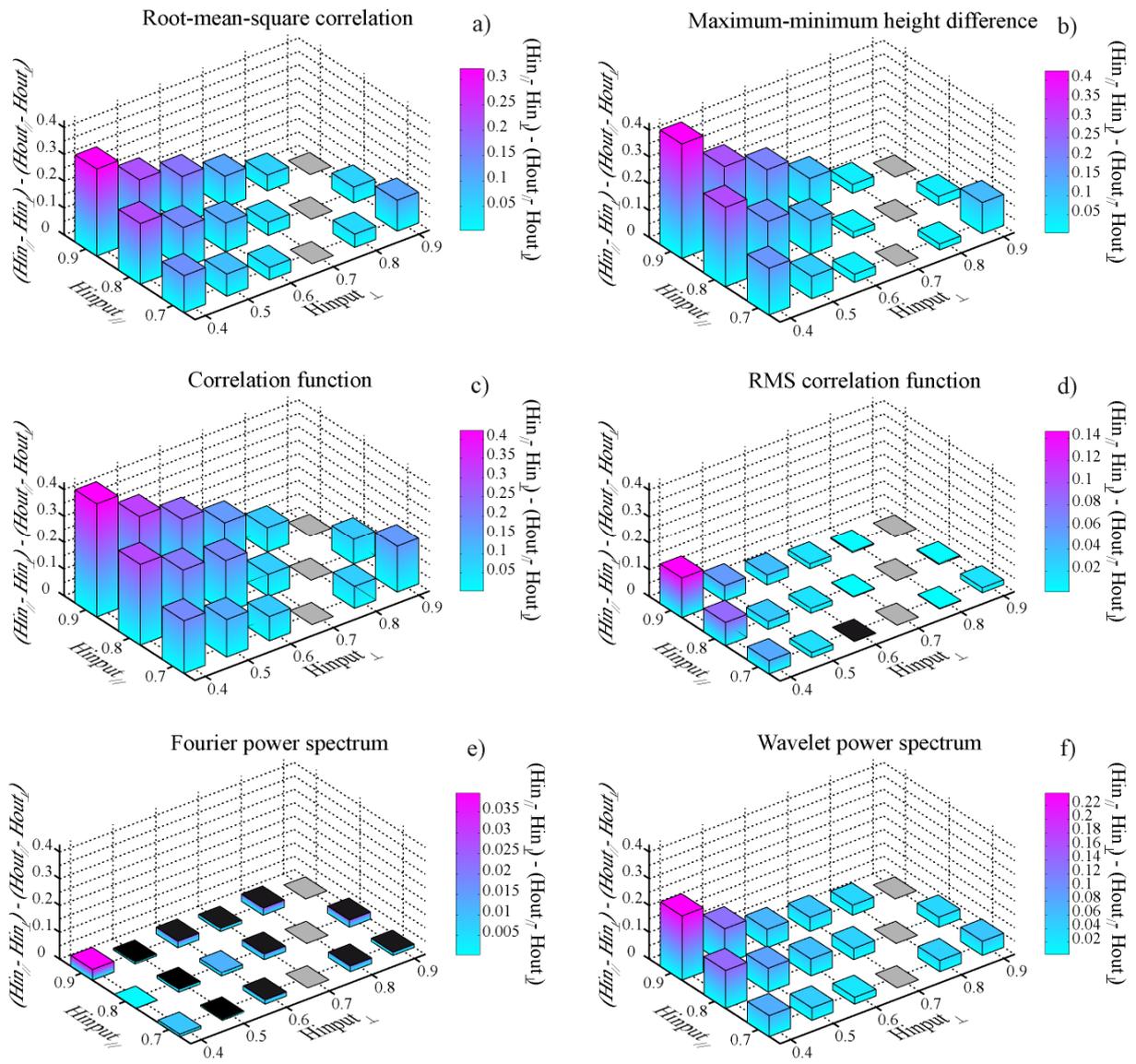

**Figure 6.**

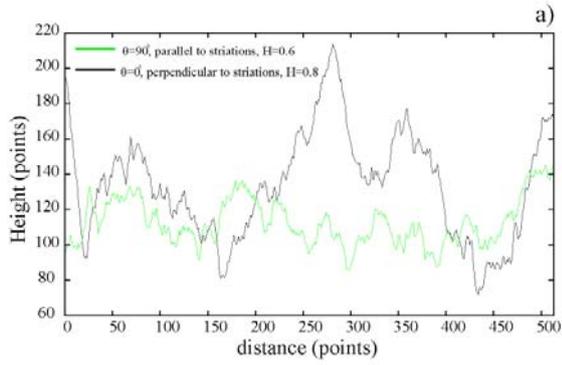
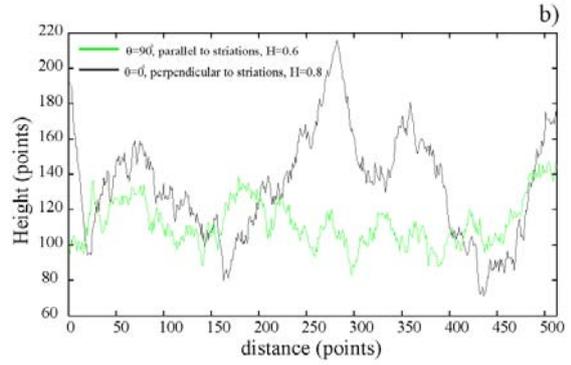
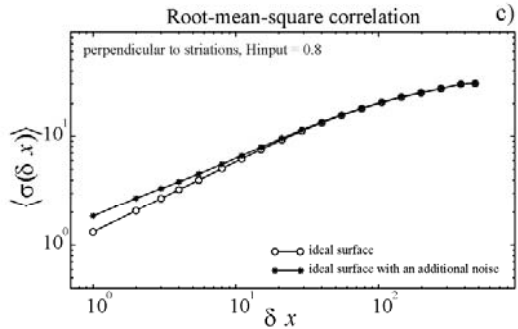
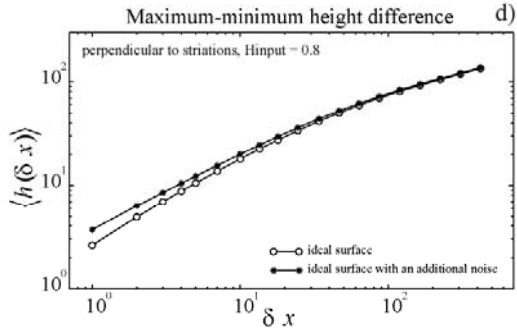
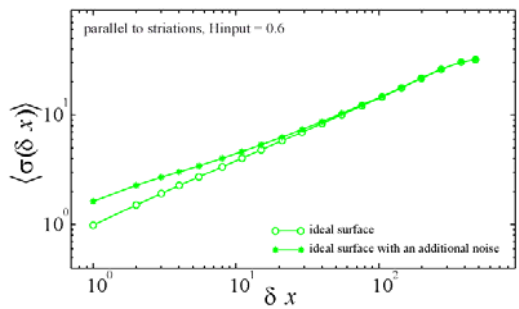
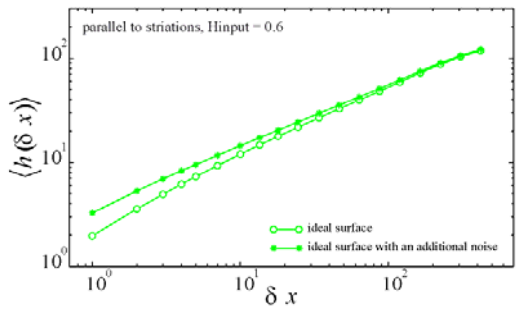
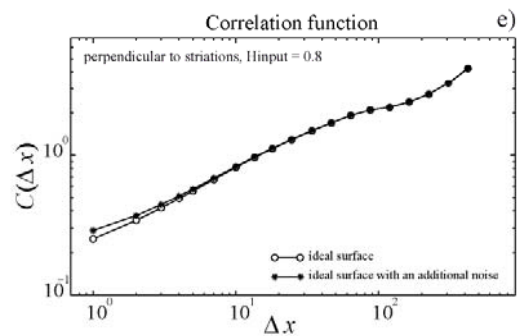
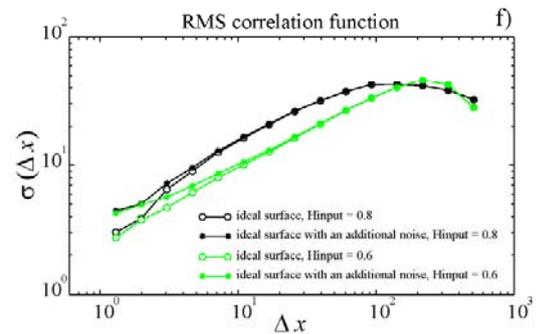
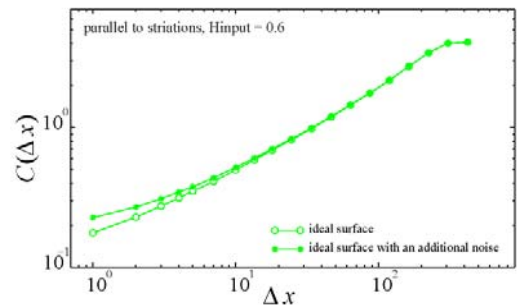
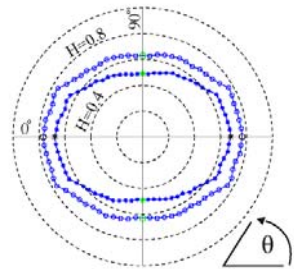

1036

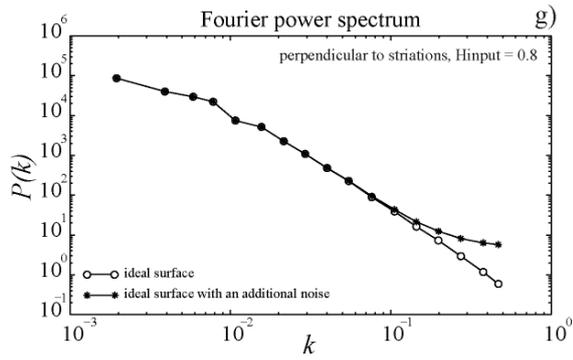
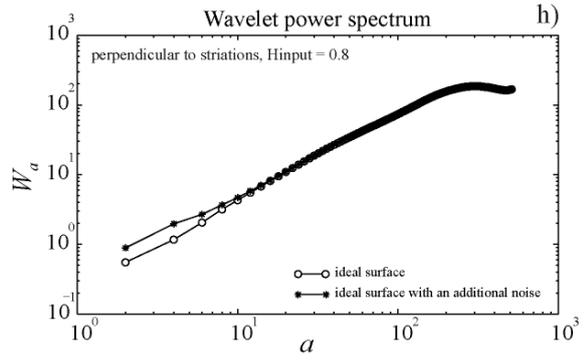
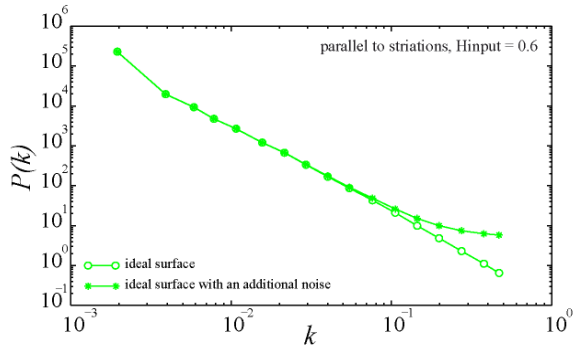
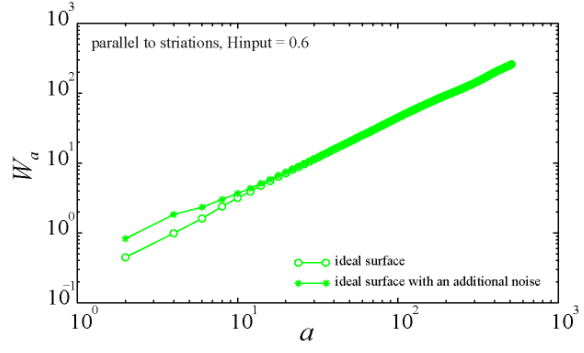

**Figure 7.**

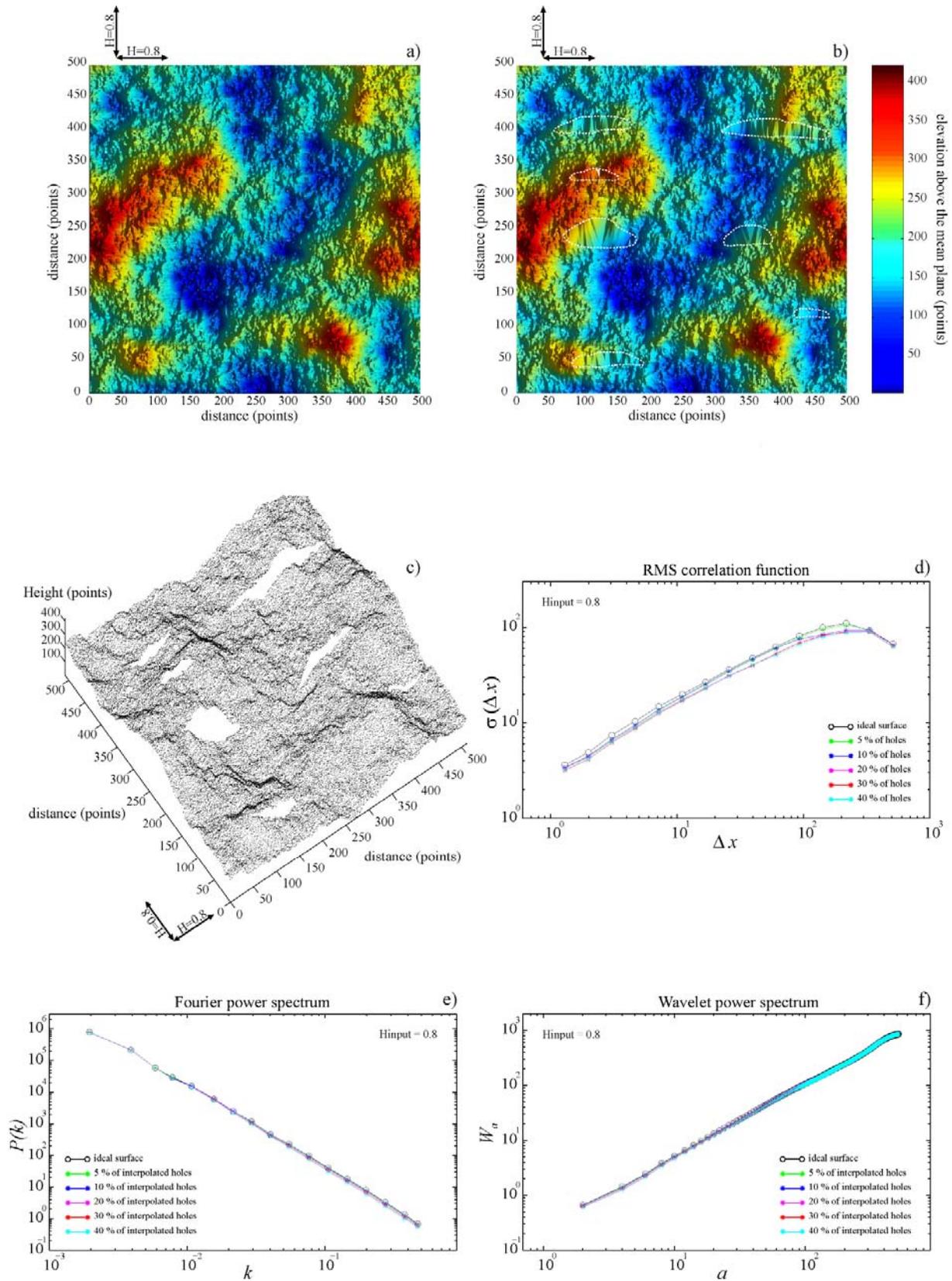

1039

1040 **Figure 8.**

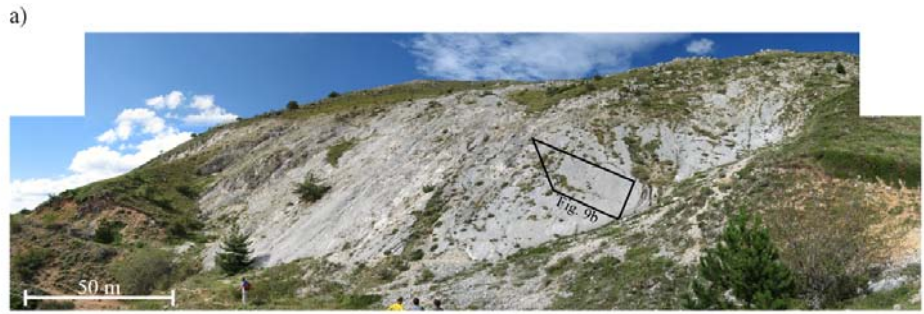
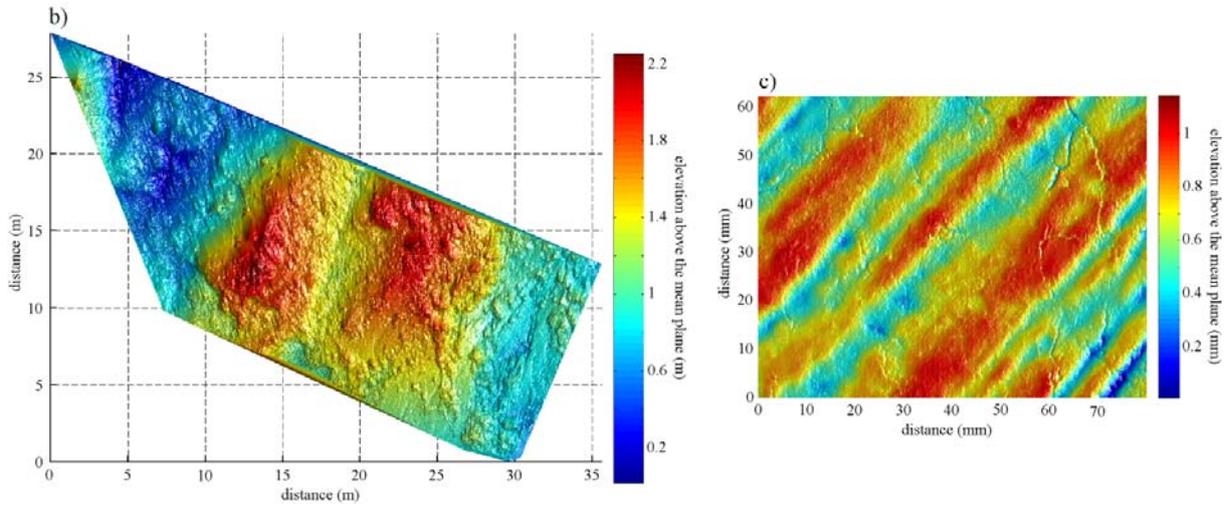

**Figure 9.**

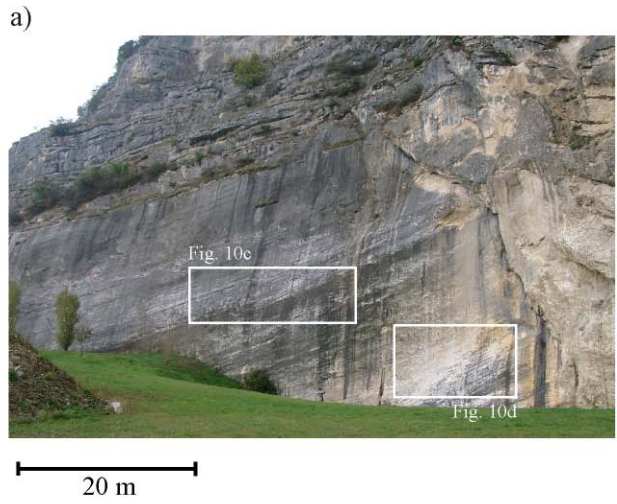
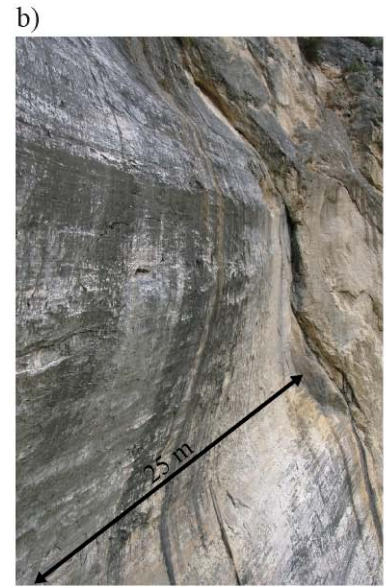
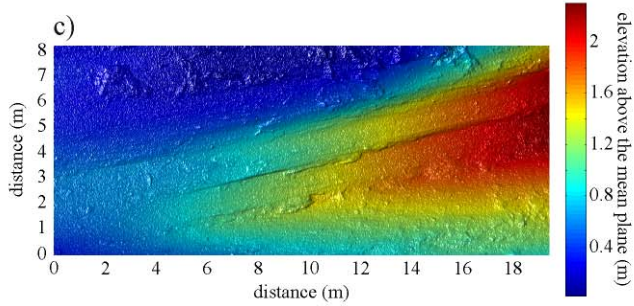
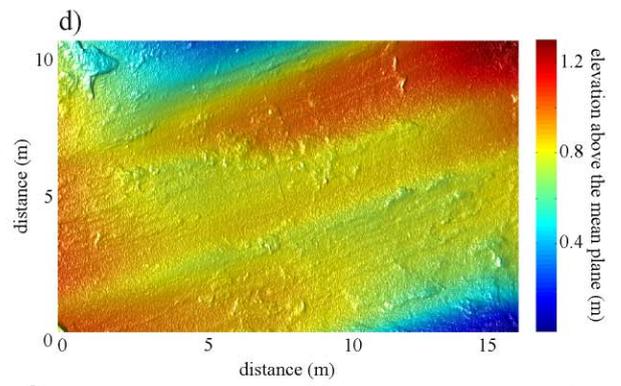
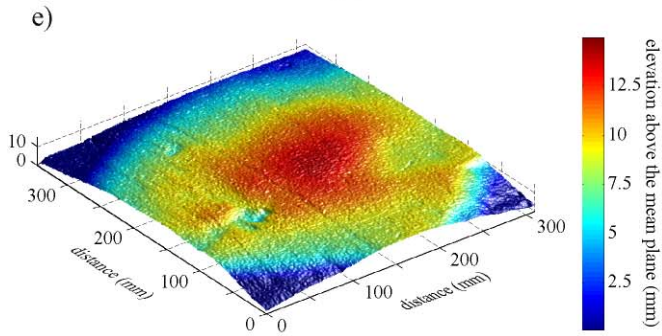
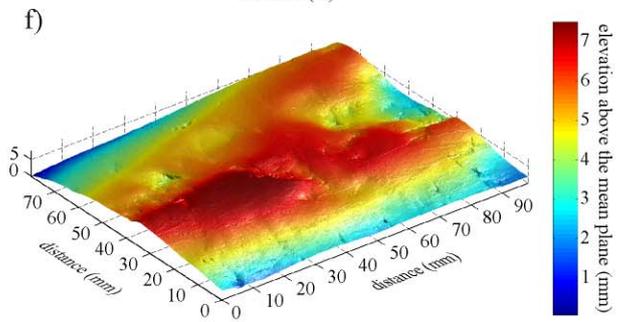

**Figure 10.**

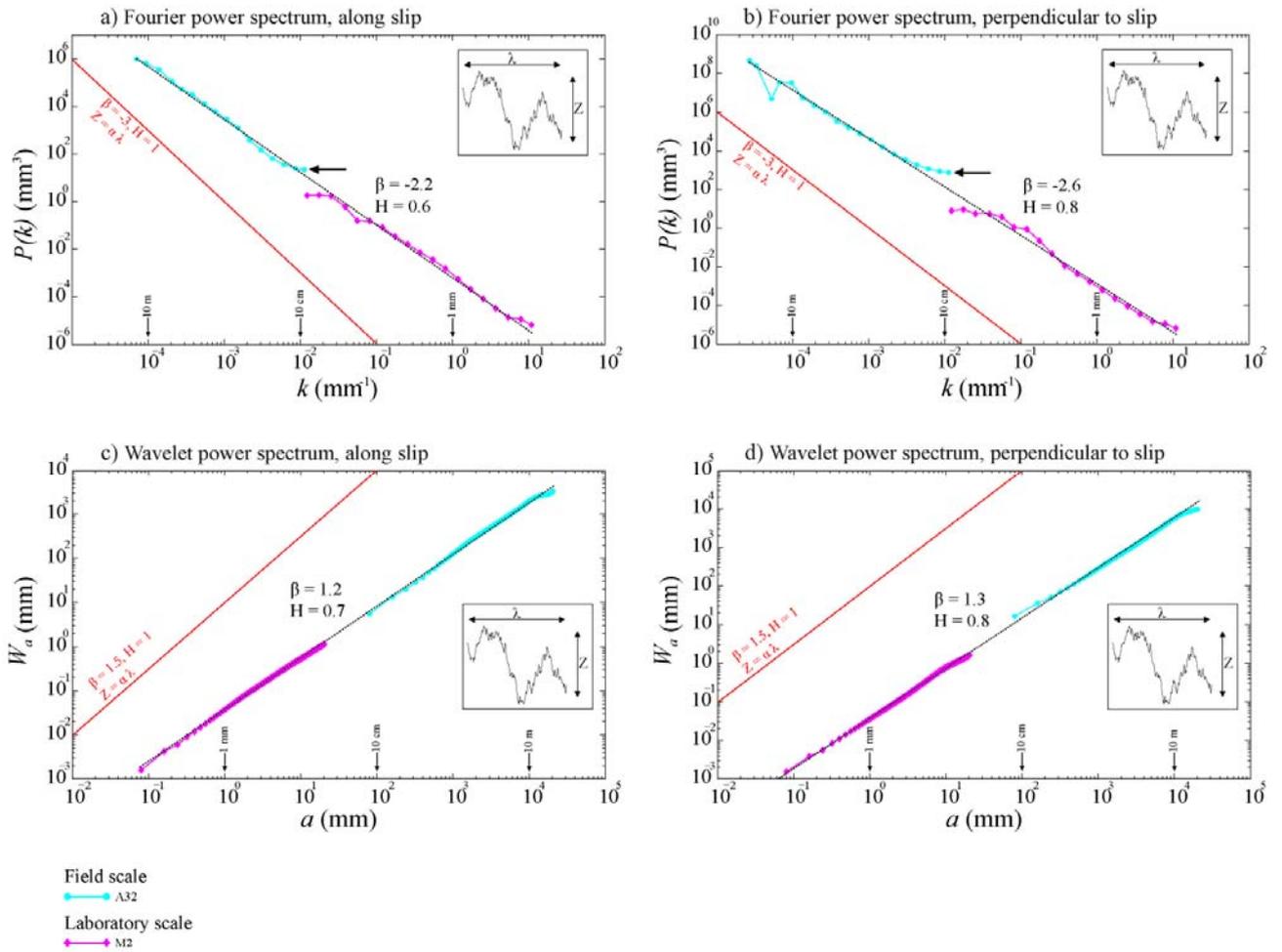

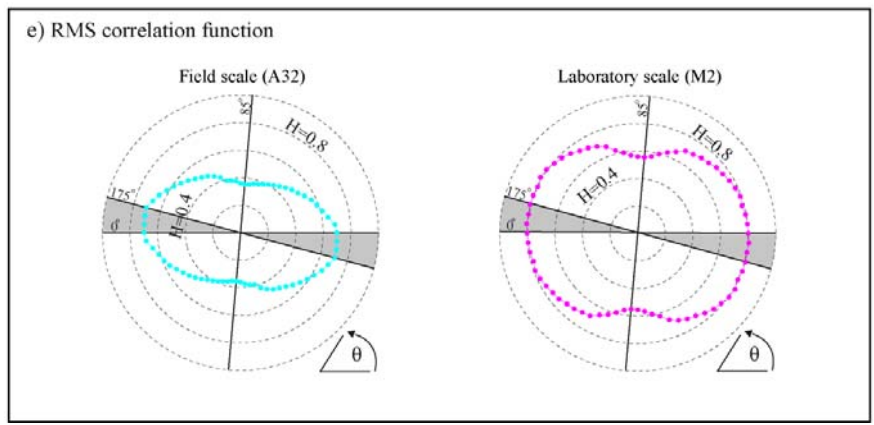

1047

1048 **Figure 11.**

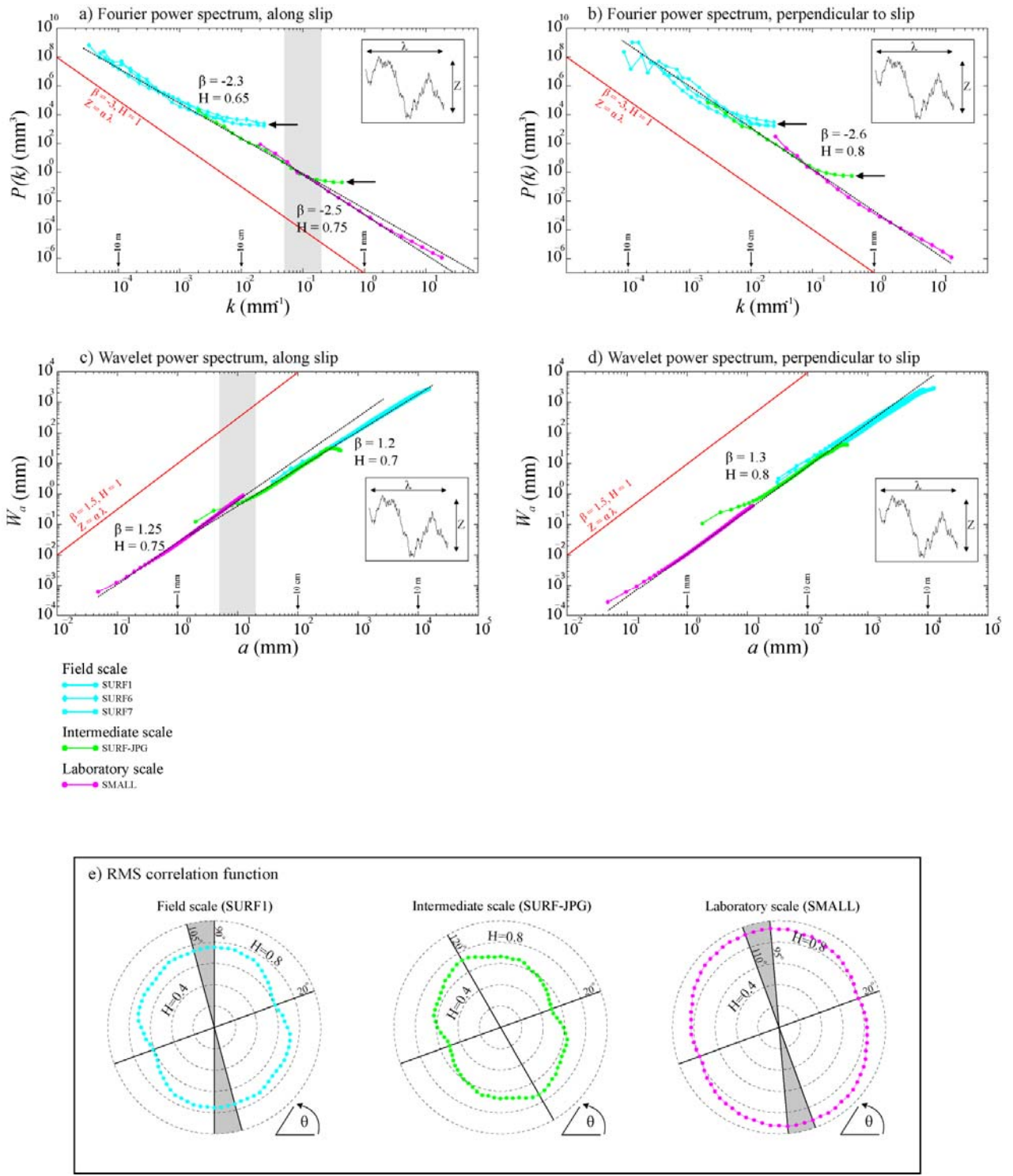

Figure 12.